\documentclass[twocolumn]{aastex631}
\usepackage{amsmath}
\usepackage{xspace}
\usepackage{amssymb}
\usepackage{xcolor}

\newcommand{\msun}{\ensuremath{M_\odot}}
\newcommand{\rsun}{\ensuremath{R_\odot}}

\newcommand{\Gaia}{\textit{Gaia}\xspace}

\begin{document}
\definecolor{darkgreen}{rgb}{0, 0.4, 0}

\title{A Sample of Extreme Eclipsing Binaries with Accretion Disks from LAMOST and ZTF}

\shorttitle{Extreme Eclipsing Binaries}
\shortauthors{Zhuang et al.}

\author[0009-0003-6191-0147]{Jiangxinxin Zhuang}
\affiliation{Department of Astronomy, Xiamen University, Xiamen, Fujian 361005, People's Republic of China} 

\author[0000-0002-2419-6875]{Zhi-Xiang Zhang}
\affiliation{Department of Astronomy, Xiamen University, Xiamen, Fujian 361005, People's Republic of China}

\author[0000-0003-3137-1851]{Wei-Min Gu}
\affiliation{Department of Astronomy, Xiamen University, Xiamen, Fujian 361005, People's Republic of China}

\author[0000-0002-7135-6632]{Senyu Qi}
\affiliation{Department of Astronomy, Xiamen University, Xiamen, Fujian 361005, People's Republic of China}

\correspondingauthor{Zhi-Xiang Zhang, Wei-Min Gu}
\email{zhangzx@xmu.edu.cn, guwm@xmu.edu.cn}

\begin{abstract}
Extreme eclipsing binaries may harbor peculiar physical properties. In this work, we aim to identify a sample of such systems by selecting binaries with pronounced eclipsing light curves, characterized of large variability ($\Delta \mathrm{mag} > 0.3$ in ZTF $g$ band) and significant differences between primary and secondary eclipses (eclipse depth ratio $>$ 20 in ZTF $g$ band). We identified 23 candidates by combining the photometric data and the LAMOST spectroscopic survey. Spectroscopic analysis revealed that all of these systems are dominated by A-type stars in the optical band. Further investigation confirmed that all 23 candidates are Algol-type binaries, with 22 of them being newly discovered. Their orbital periods range from 2.57 to 19.21 days. These systems consist of low-luminosity, highly stripped subgiant donors and accreting A-type stars. The donor stars, with radii of 2.5--8.9\,$R_\odot$ and effective temperatures around 4000\,K, have typical masses of \(M_2 \sim 0.3\,M_\odot\), indicating substantial mass loss through Roche-lobe overflow. The presence of ellipsoidal variability and H$\alpha$ emission provides strong evidence for ongoing mass transfer. By fitting the spectral energy distributions, spectra, and light curves, we found that most of the accretors have luminosities lower than expected from the mass-luminosity relation, aligning with the predicted faint phase for mass-gaining stars. Three objects of our sample exhibit pulsations with periods from 18 minutes to 8 hours, providing opportunities for asteroseismic studies. The low mass transfer rates and stability make the sample excellent systems for studying mass accretion, advancing our understanding of the Algol-type binary evolution.
\end{abstract}


\section{Introduction} \label{sec:intro}
High-precision photometric and spectroscopic time-series measurements allow eclipsing binaries (EBs) to serve as prime targets for obtaining accurate fundamental stellar parameters, as the mass and radius of each component in EBs can be determined with exceptional precision to within 1 percent \citep{clausen2008absolute}. 

The Large Sky Area Multi-Object Fiber Spectroscopic Telescope \citep[LAMOST;][]{cui2012large} spectroscopic survey has accumulated a substantial amount of spectral data, while the Zwicky Transient Facility \citep[ZTF;][]{2019PASP..131a8003M} survey provides complementary time-series photometry. The integration of these two datasets can enhance the discovery and study of eclipsing binaries through both spectroscopy and photometry. Additionally, {\it Gaia} DR3 \citep{vallenari2023gaia} provides space-based astrometric distance measurements, which can simultaneously constrain the fundamental physical parameters of the systems through a joint fitting of observational data.

Partial of the eclipsing binaries exhibit extreme photometric characteristics, including pronounced brightness variations \citep{2024A&A...686A.199K} and dramatic eclipses \citep{2015ApJ...808..179D,2022A&A...664A..96Z}, as well as unique spectroscopic features that distinguish them from ordinary eclipsing binaries. The extraordinary 
observational features of these systems often indicate extreme physical characteristics, including extreme mass ratios \citep{2024AJ....168...50L}, extreme temperature ratios \citep{2019ApJ...884..126M}, and intense mass transfer \citep{2025arXiv250216059L}. 

Therefore, it is essential to obtain a sample with unique physical properties by focusing on specific extreme observational features. Specifically for this work, the selection criteria we adopt are as follows: (1) extreme differences in the depths of the two eclipses, (2) exceptionally large variability amplitudes, (3) the presence of emission lines in their spectra. These criteria are designed to identify binaries undergoing mass transfer, where the two companion stars exhibit a large temperature difference. We obtained a sample based on these standards and 
our joint spectral analysis of the selected binary systems demonstrates striking uniformity in their fundamental characteristics. Each system within the sample comprises an A-type primary star accompanied by a cooler subgiant companion. These binaries exhibit completed mass ratio reversal while maintaining ongoing mass transfer, fulfilling the defining criteria of classical Algol-type systems.
 
Algol-type binaries, which are close, usually eclipsing, semi-detached binary star systems, represent a unique subset. Typically, Algols consist of a more luminous and massive late B-, A-, or F-type star alongside a less luminous and less massive evolved giant or subgiant companion\citep{kopal1959close}. Most of the less massive stars are overflowing their Roche lobe and losing mass to the accretors. The evolution of Algol-type binaries is particularly special because the less massive stars appear more evolved compared to their more massive counterparts. This phenomenon is generally attributed to mass transfer between the two stars.
The properties of stars in Algol-type binaries are used to calibrate theoretical stellar evolution model \citep{2011A&A...528A..16V,2015ASPC..496..137F}, support asteroseismology \citep{2022Ap&SS.367...22U}, and research post-mass-transfer stars, including blue straggler stars \citep{chen2009primordial}, extremely low mass (ELM) white dwarfs or pre-ELM \citep{pelisoli2019gaia}, and barium stars \citep{2021MNRASbarium}. Meanwhile, Algol binaries serve as a valuable laboratory for studying various types of accretion disks \citep{1989Spectroscopic,van2016accretion}. Furthermore, Algol-type binaries may also evolve into compact white dwarf binaries, which are considered important targets for future gravitational wave detection by the LISA space observatory \citep{lu2020possible,burdge2023orbital}. 

Among the 23 binary systems reported in this study, 22 are newly identified, while one, J1905, was recently specifically analyzed by \citet{zhang2024photometric}. These systems exhibit several characteristics typical of Algol-type binaries, including a reversed mass ratio and ongoing mass transfer. The correlation of the emission line profile and variability with orbital periods are consistent with previously reported Algol-type binaries \citep{richards1999morphologies}. Additionally, three of our objects display significant pulsation signals, which also reported in many previous works on Algol-type binaries \citep{pigulski2007pulsating, gillet2019dynamical}. However, despite these similarities, our sample also presents unique observational characteristics. Notably, these binaries occupy a more constrained parameter space, with donor stars that have significantly lower masses than those in most previously reported Algols. Furthermore, the donor stars in our sample appear more bloated and cooler, suggesting that they are in later evolution stage compared to the most known Algol-type binaries.
 
Binary systems selected based on simple yet extreme photometric variability exhibit remarkable consistency in their physical parameters. This strong association suggests a deep intrinsic link between observable characteristics and the underlying properties of these systems \citep{2024A&A...686A.199K}. Moreover, it provides valuable insight into the population distribution of detectable binary stars. By refining systematic selection criteria, we can further explore the connection between evolutionary stages and observational signatures, offering new constraints for theoretical modeling and advancing our understanding of binary system evolution \citep{2015ASPC..496..137F}.

In this paper, we describe the selection of our sample based on photometric and spectral observations, 
as detailed in Section \ref{sec:sample-data}. Stellar parameters of the stars are derived through the 
fitting of SEDs, spectra, and light curves, discussed in Sections \ref{subsec:sed}, 
\ref{subsec:spec-fit}, and \ref{subsec:lc-fit}, respectively.
The masses of the donor stars are estimated in Section \ref{subsec:mass_const}.
The Color-Magnitude Diagram (CMD) is shown in Section \ref{subsec:CMD}, while Section \ref{subsec:ha} 
offers a detailed spectroscopic analysis, focusing on the H$\alpha$ emission lines and the presence 
of accretion disks. Pulsation characteristics are discussed in Section \ref{subsec:pulsation}.
Finally, the discussion and summary of our findings are provided in Sections \ref{sec:dis} 
and \ref{sec:summary}.

\section{Sample selection and data reduction} \label{sec:sample-data}
We design to select binary systems with extreme light curves, including large eclipse depth differences and variability (similar to objects like J0935 reported by \citealt{2022Mu}). We aim to investigate the physical properties of binaries selected under these extreme conditions. Additionally, we impose an extra selection criterion requiring the presence of emission lines in the spectra. Such features may indicate intriguing physical properties, such as obvious stellar activity, ongoing mass transfer, or irradiation effects.

\begin{figure*}
\centering 
\includegraphics[width=0.9\textwidth]{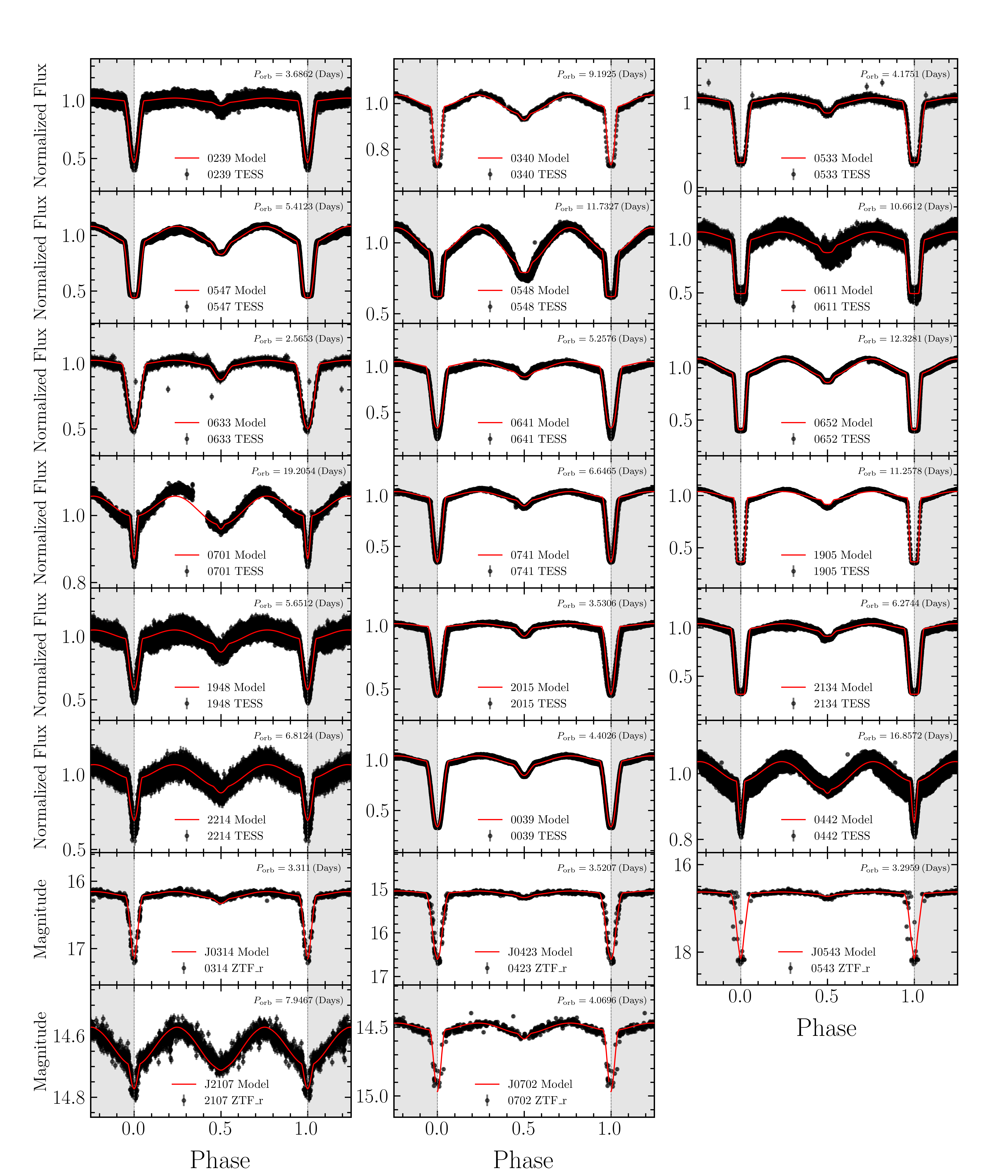}
\caption{Light curves of 23 binaries folded by their photometric periods. The photometric data were sourced from \textit{TESS} or ZTF r band. Five of these binaries have nearby stars that significantly affect the \textit{TESS} data. To minimize the impact of third light on the light curves of these binaries, we utilized data from the ZTF survey. The red line represents the light curve model resulting from \texttt{ellc} fitting. The period of each source is shown in the upper right corner.}
\label{fig:lc}
\end{figure*}

\citet{xu2022catalog} identified a catalog of variable sources from LAMOST and ZTF by using a statistical modeling approach. We utilize this catalog to select candidates. We first request the objects to have high signal-to-noise ratios (SNR \(> 10\)) in LAMOST observations. For the remaining objects, we downloaded their ZTF-$g$ band light curves and used the Lomb-Scargle periodogram \citep{1976Least} implemented in the \texttt{astropy} \citep{2013astropy, 2018astropy} package to determine their photometric periods. These periods are subsequently used to phase-fold the light curves. We select phase-folded light curves that show a significant difference in eclipse depths and noticeable ellipsoidal modulation. These ZTF-$g$ band light curves must meet the following criteria: the variability amplitude exceeds 0.3 magnitudes, and the eclipse depth ratio (primary to secondary) is greater than 20. 
If the photometric data from the ZTF eclipse phase is insufficient for clear identification, we will collect additional photometric data from the Transiting Exoplanet Survey Satellite \citep[TESS;][]{ricker2015transiting} survey. 
The significant difference in the depths of the primary and secondary eclipses is mainly due to the substantial surface temperature difference between the two stars in the binary system, as illustrated in Figure~\ref{fig:lc}. 

We focus on binary systems that may undergo interactions. To identify these, we check the LAMOST spectra, highlighting on sources with H$\alpha$ emission through visual inspection. This criterion helps us identify systems with potential mass transfer, as demonstrated by the H$\alpha$ emission in their LAMOST spectra, shown in Figure~\ref{fig:ha}.

Ultimately, we have identified a sample of 23 binaries with marked difference in eclipse depths and H$\alpha$ emission. Subsequent research (Section \ref{subsec:fit-result}) indicates that all these systems are Algol-type binaries with ongoing mass transfer process. 
We will refer to the hotter component in each binary as the ``accretor'' and the cooler component as the ``donor.'' The basic information about these binaries is presented in Table \ref{table:1}.
We emphasize that our sample selection process was independent of any preconceived type of the systems, focusing solely on observational characteristics. The resulting sample, unexpectedly, consists entirely of Algol binaries dominated by A-type stars in the optical band. 

\section{Analysis and Results} \label{sec:analysis}

\subsection{Spectral energy distributions} \label{subsec:sed}

\begin{figure*}
\centering
\includegraphics[width=0.98\textwidth]{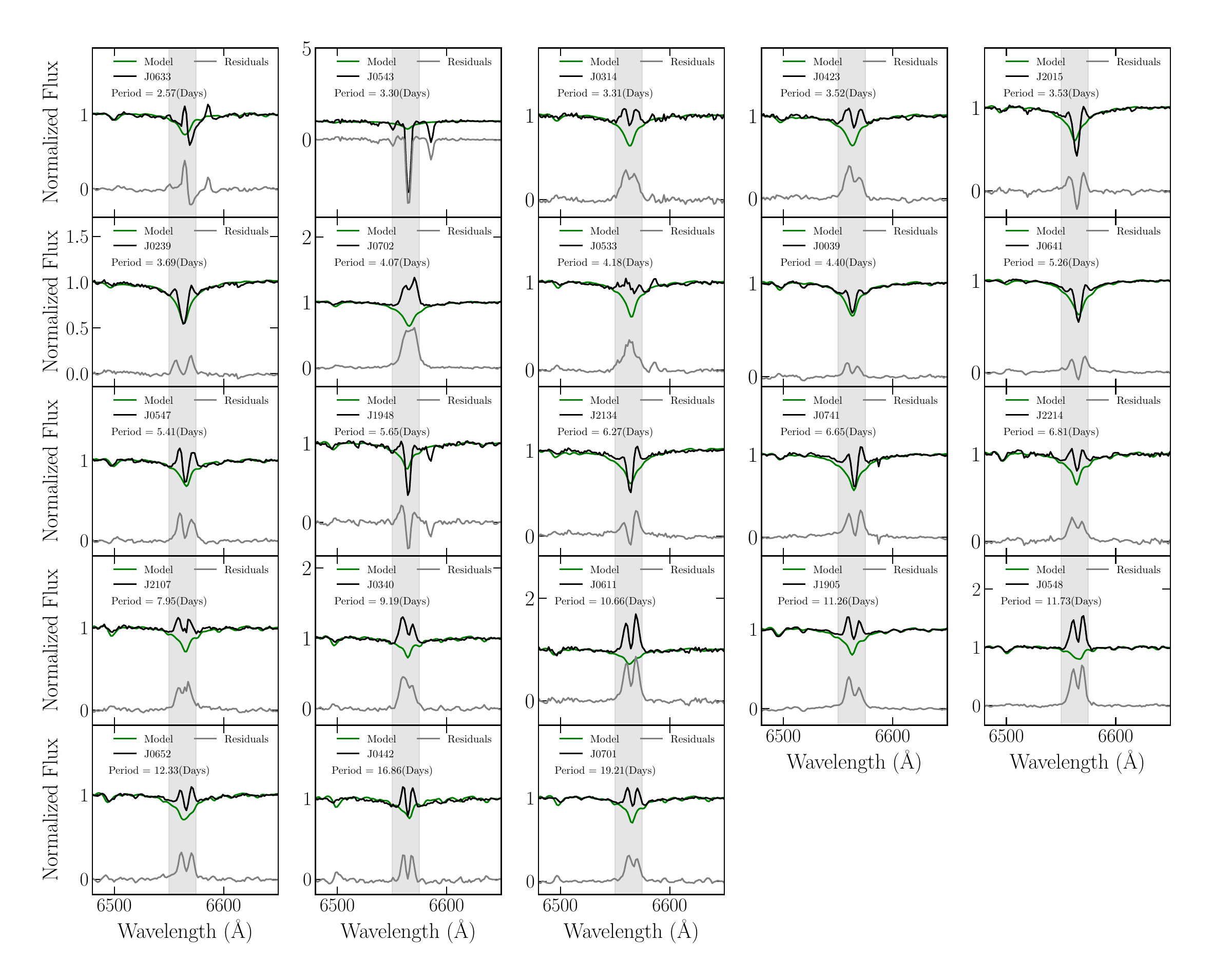}
\caption{Zoom-in in the vicinity of H$\alpha$ emissions from LAMOST spectra. The black lines represent the normalized spectral flux, while the green line shows their spectral models derived from the joint fit. The grey curve shows the residuals (observation$-$template) from this fit. The subgraphs are arranged in the order of the systems' periods.
\label{fig:ha}}
\end{figure*}

\figsetstart
\figsetnum{3}
\figsettitle{Results of the SED, spectrum, and light curve fitting.}

\figsetgrpstart
\figsetgrpnum{3.1}
\figsetgrptitle{J0039
}
\figsetplot{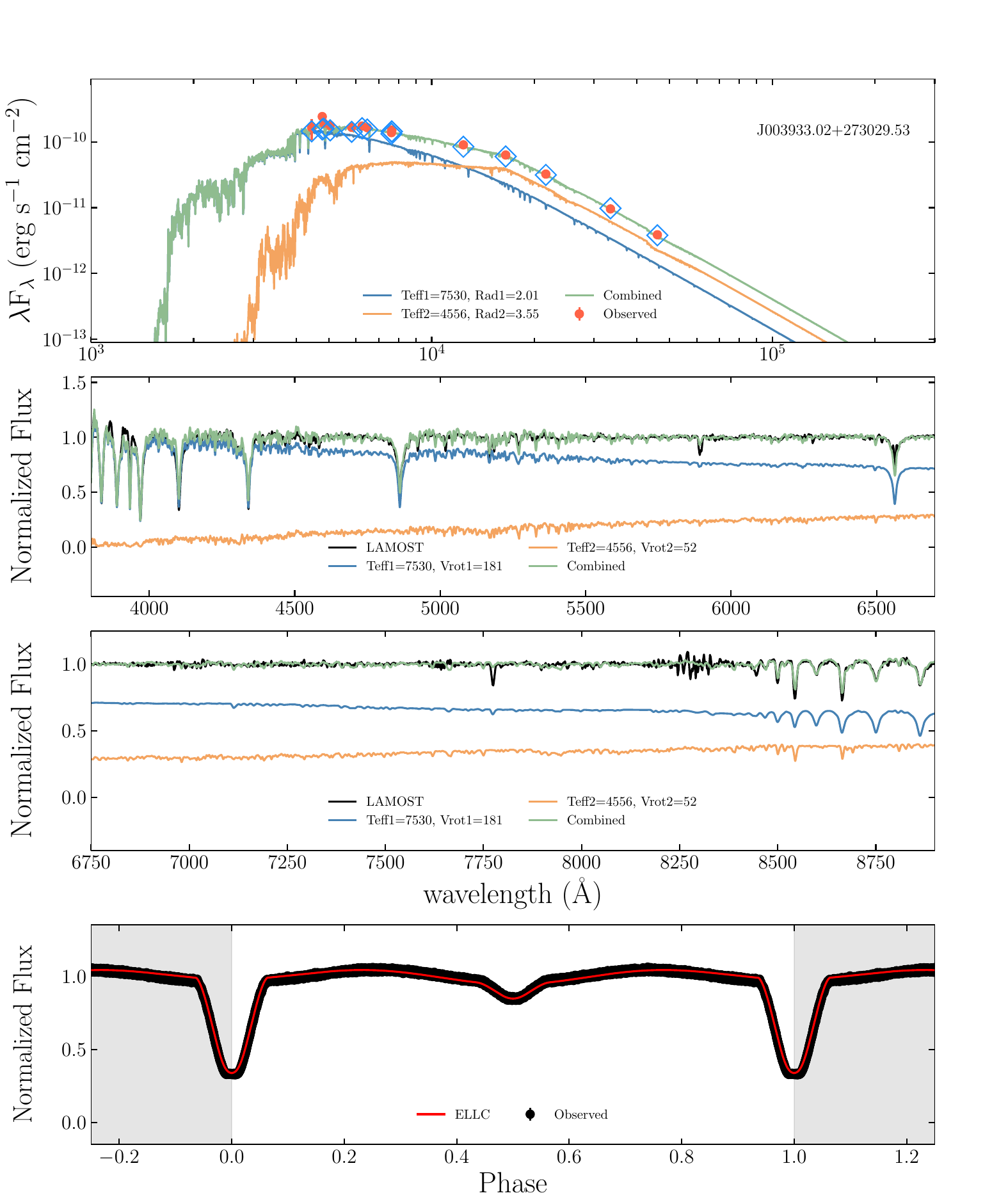}
\figsetgrpnote{ }
\figsetgrpend

\figsetgrpstart
\figsetgrpnum{3.2}
\figsetgrptitle{J0239
}
\figsetplot{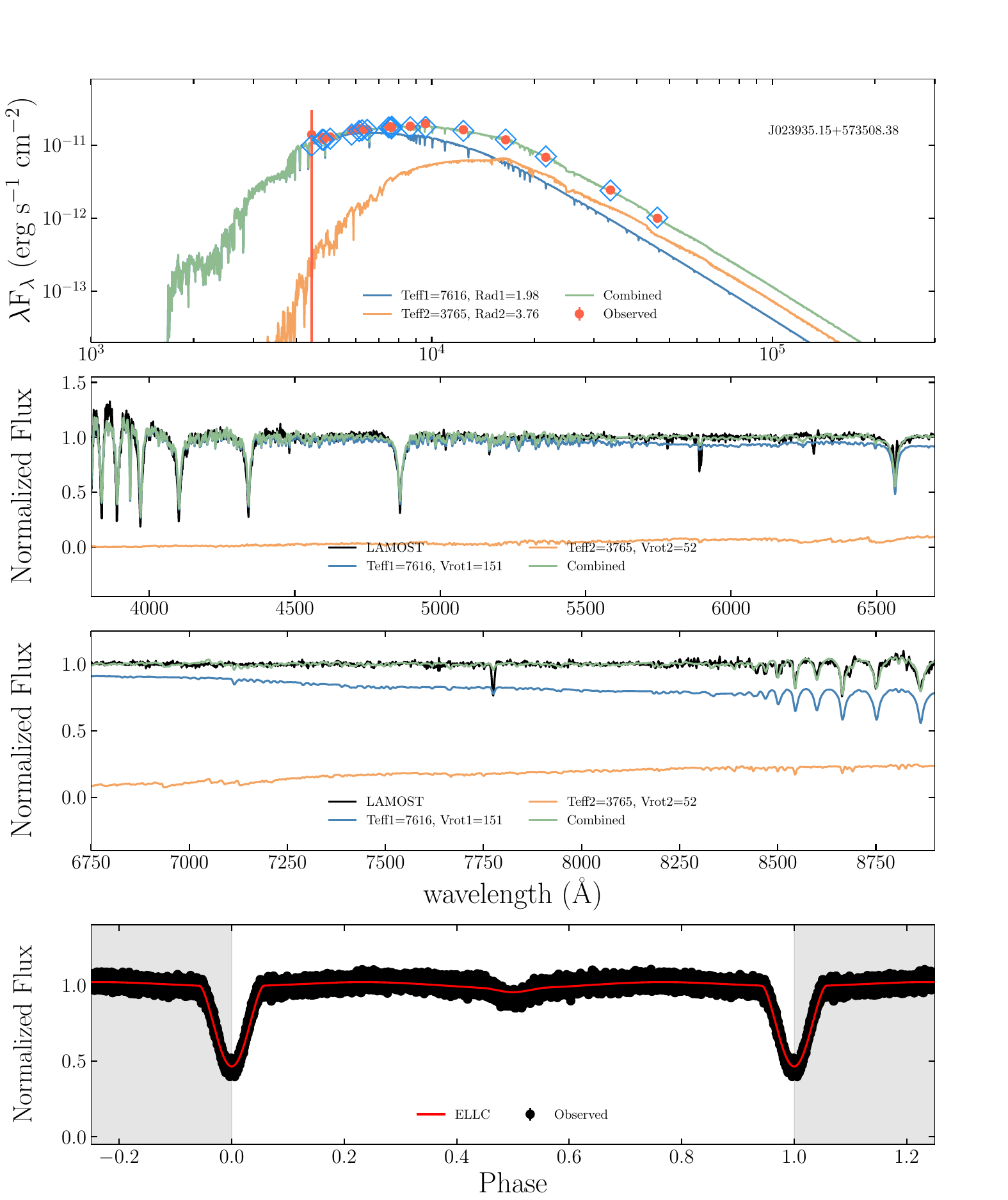}
\figsetgrpnote{ }
\figsetgrpend

\figsetgrpstart
\figsetgrpnum{3.3}
\figsetgrptitle{J0314
}
\figsetplot{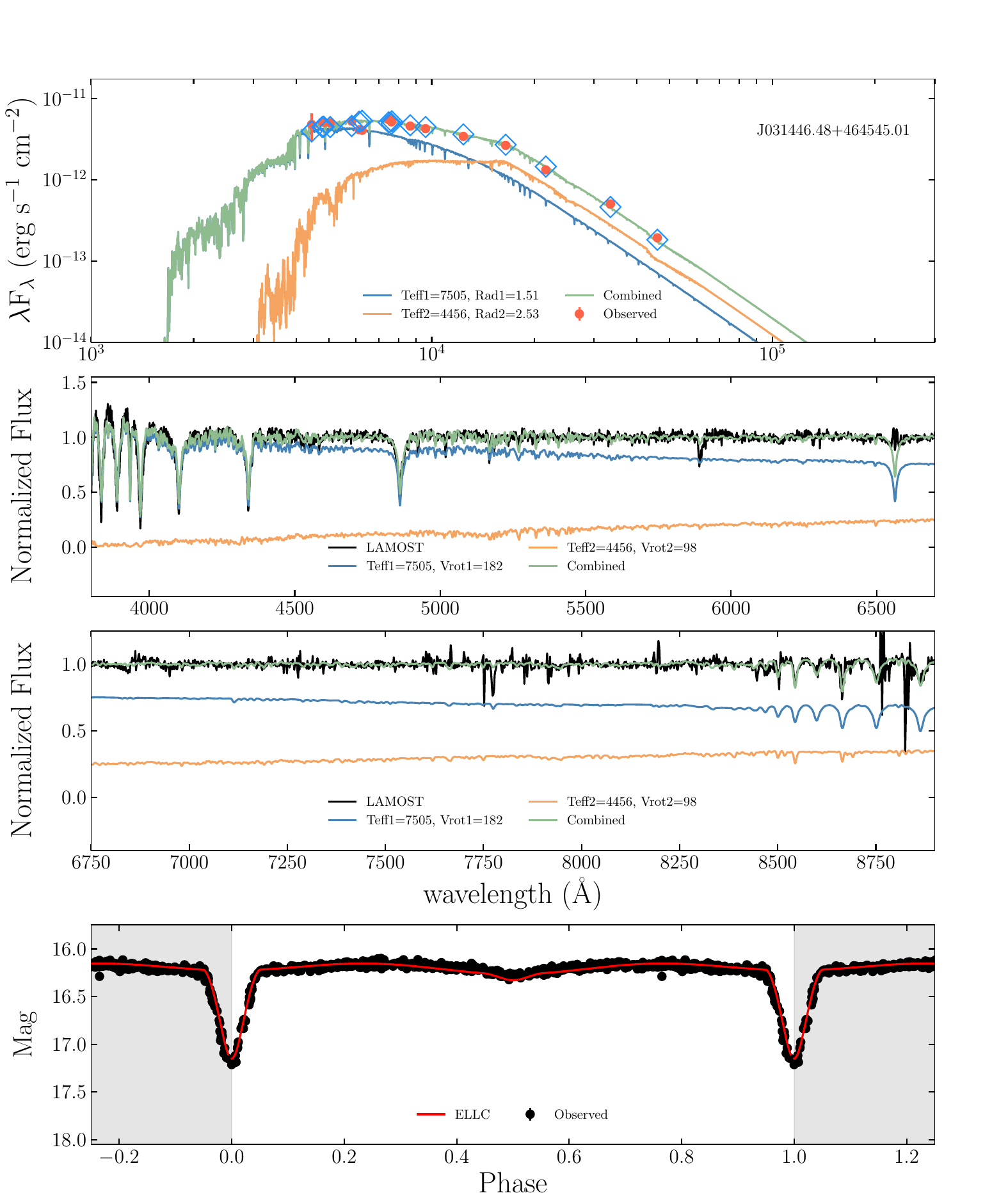}
\figsetgrpnote{ }
\figsetgrpend

\figsetgrpstart
\figsetgrpnum{3.4}
\figsetgrptitle{J0340
}
\figsetplot{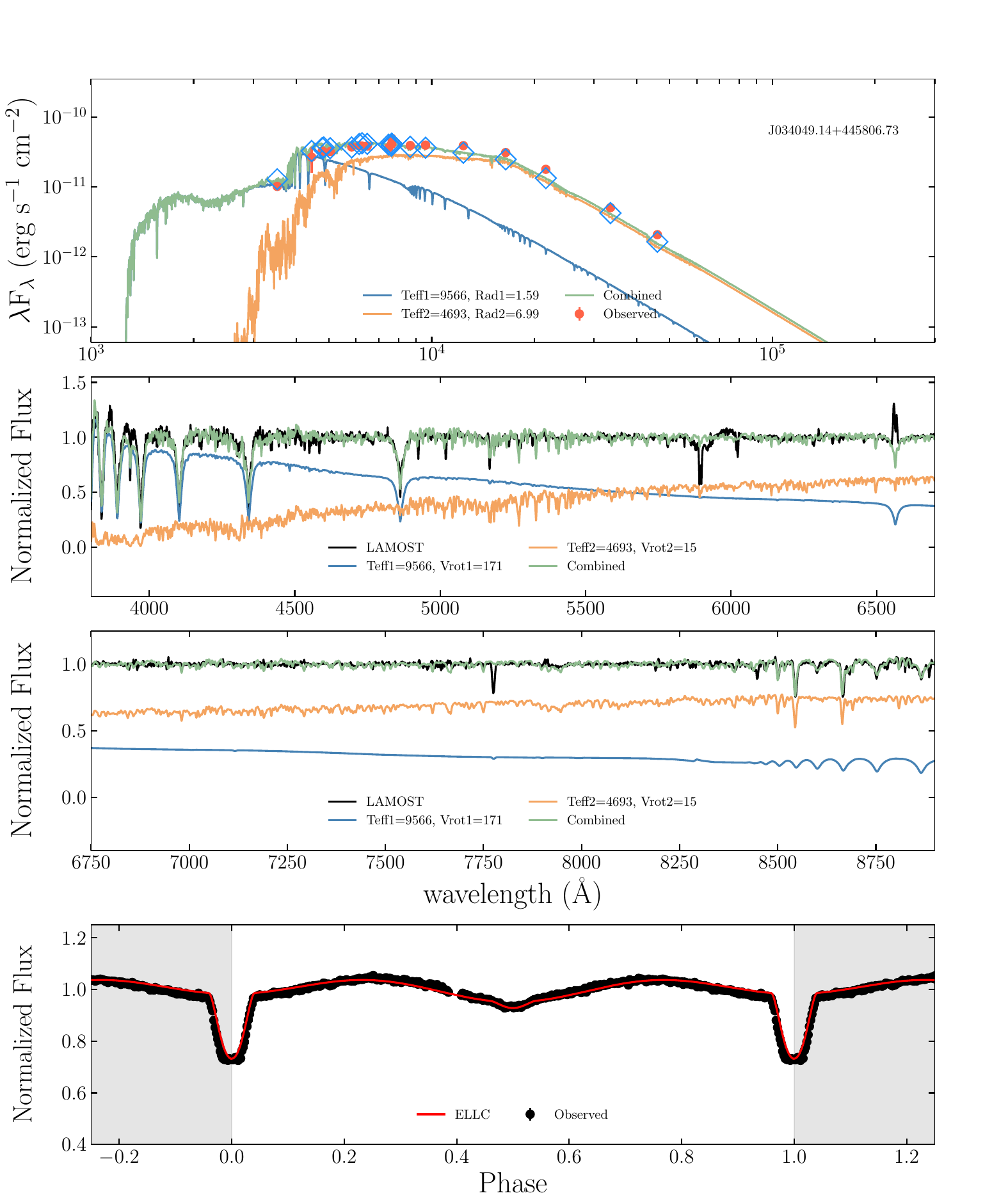}
\figsetgrpnote{ }
\figsetgrpend

\figsetgrpstart
\figsetgrpnum{3.5}
\figsetgrptitle{J0423
}
\figsetplot{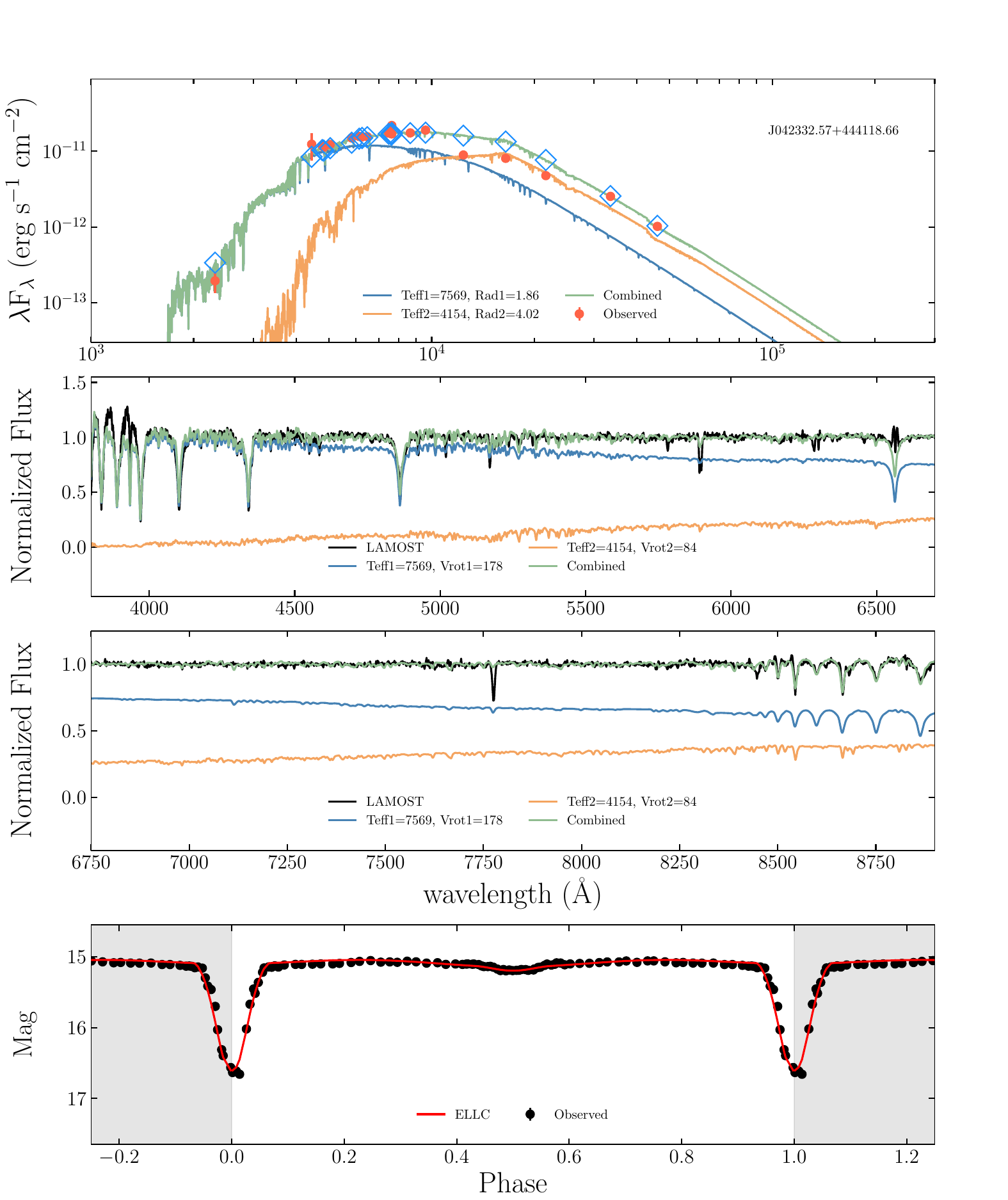}
\figsetgrpnote{ }
\figsetgrpend

\figsetgrpstart
\figsetgrpnum{3.6}
\figsetgrptitle{J0442
}
\figsetplot{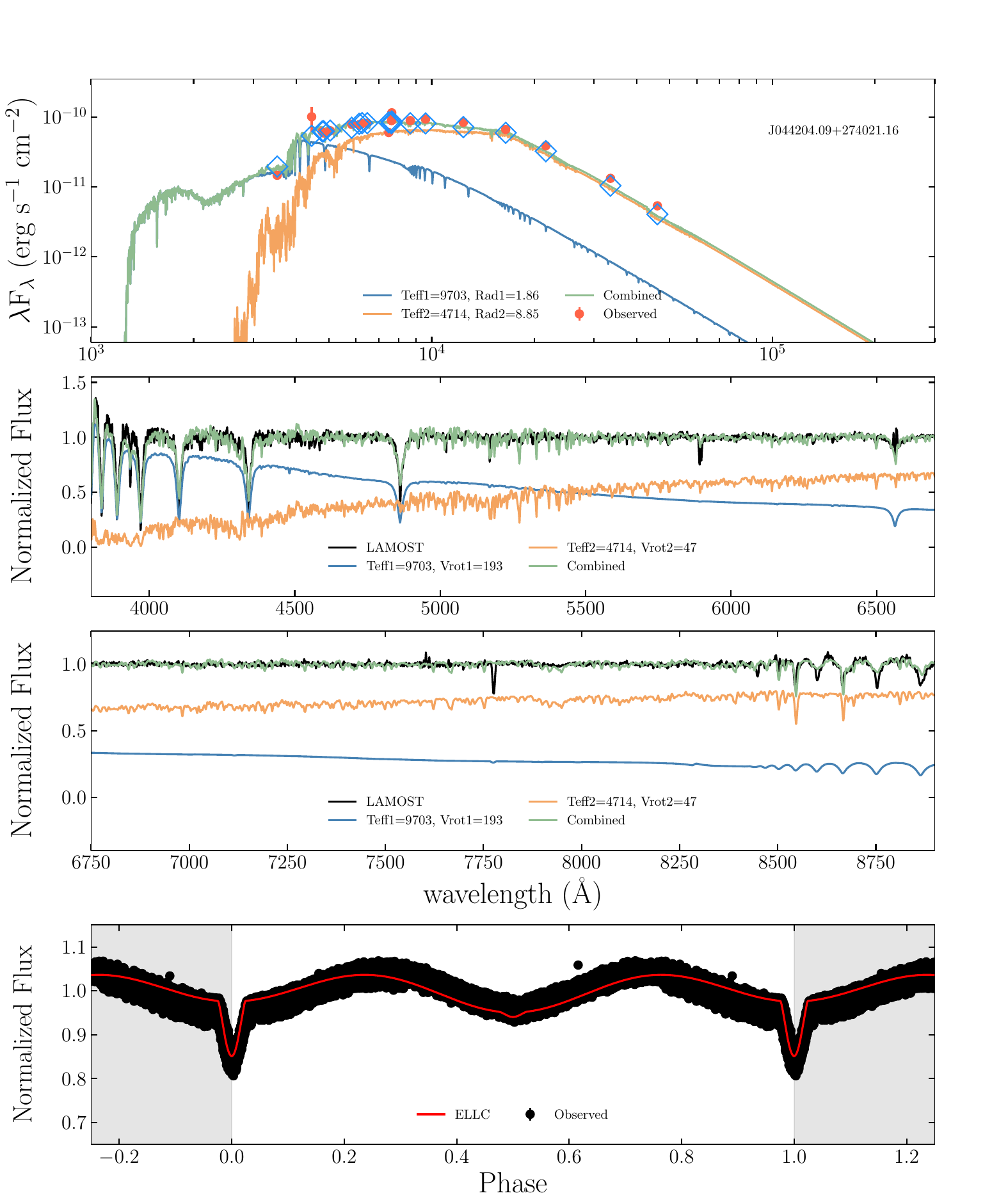}
\figsetgrpnote{ }
\figsetgrpend

\figsetgrpstart
\figsetgrpnum{3.7}
\figsetgrptitle{J0543
}
\figsetplot{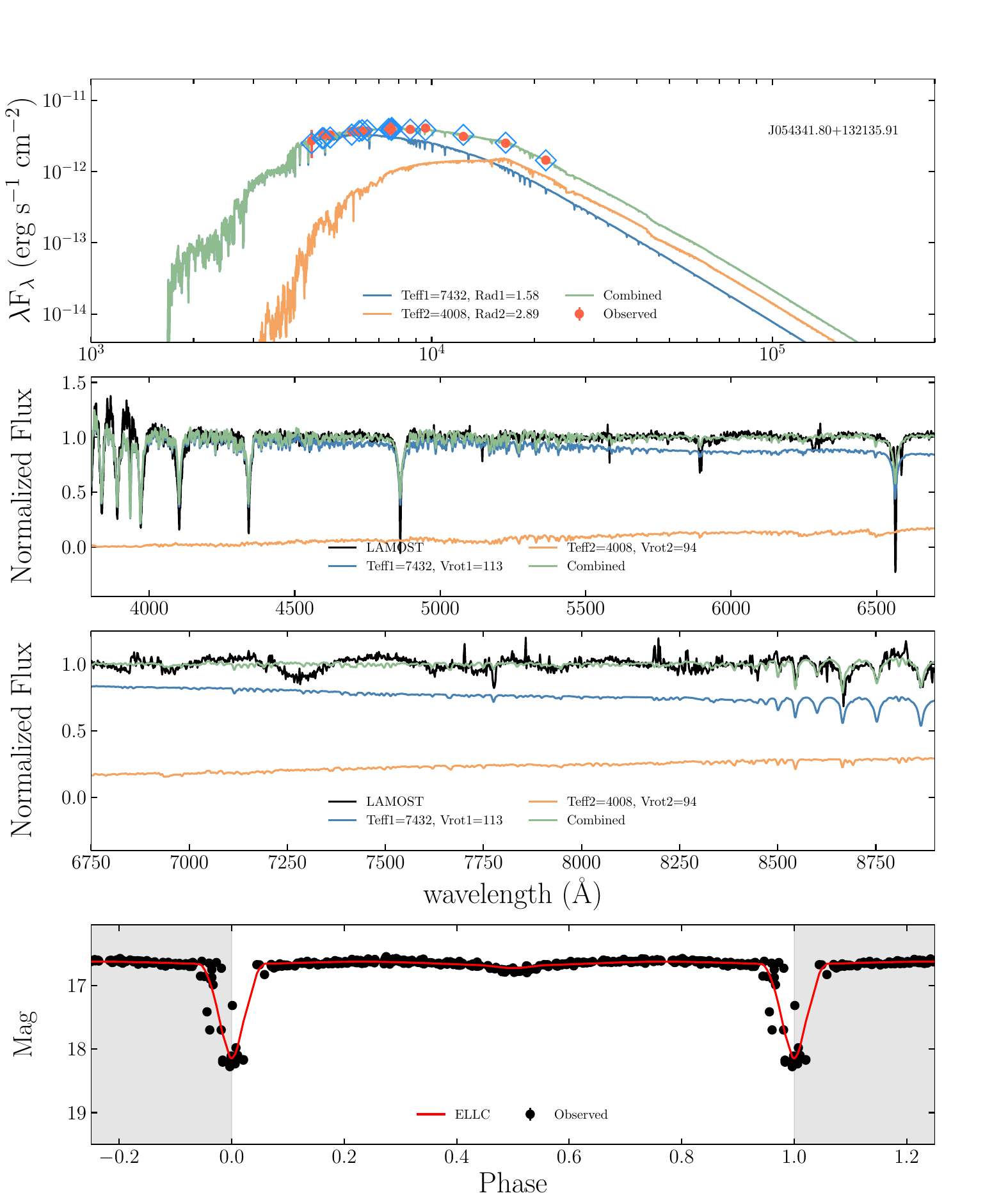}
\figsetgrpnote{ }
\figsetgrpend

\figsetgrpstart
\figsetgrpnum{3.8}
\figsetgrptitle{J0547
}
\figsetplot{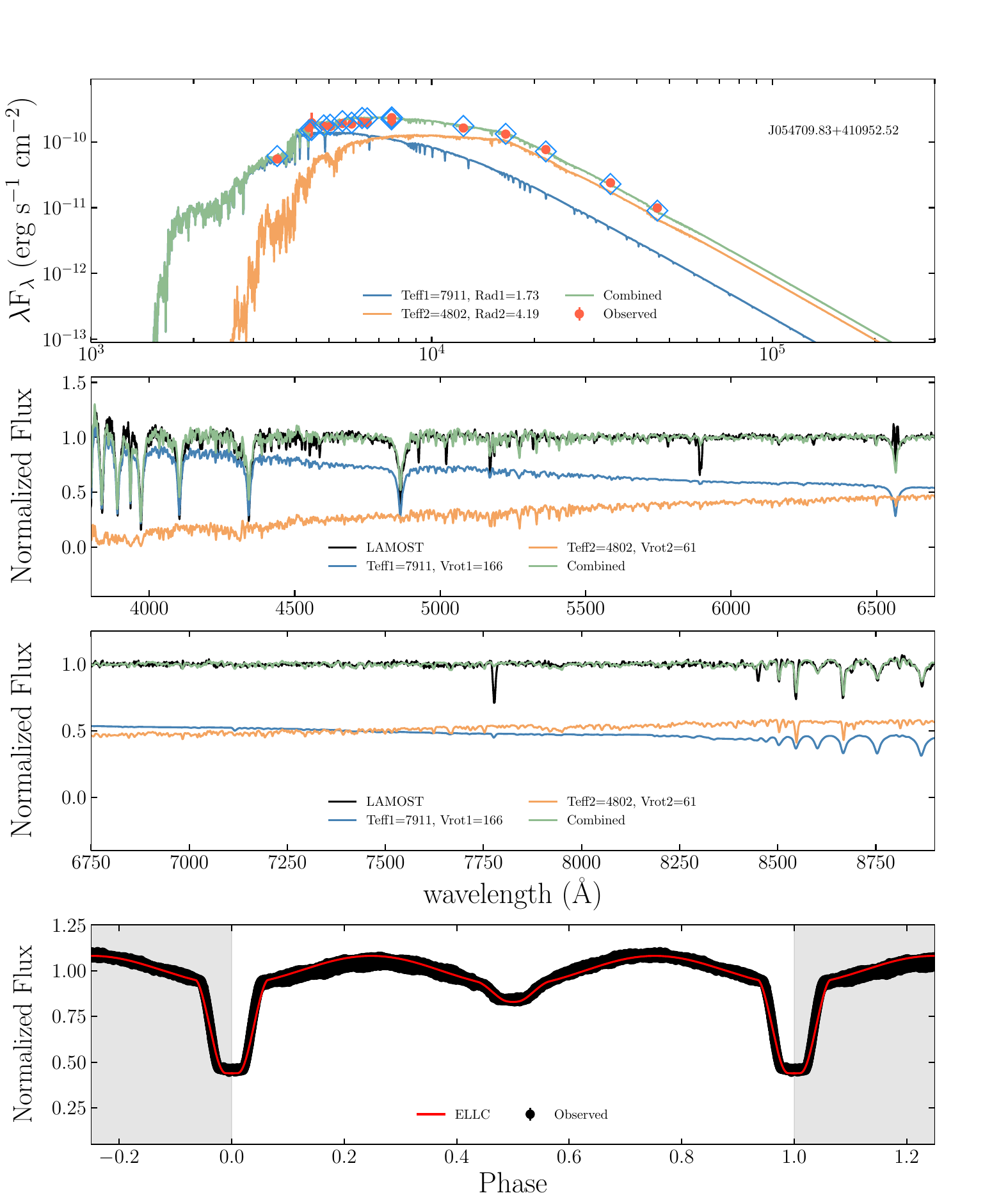}
\figsetgrpnote{ }
\figsetgrpend

\figsetgrpstart
\figsetgrpnum{3.9}
\figsetgrptitle{J0548
}
\figsetplot{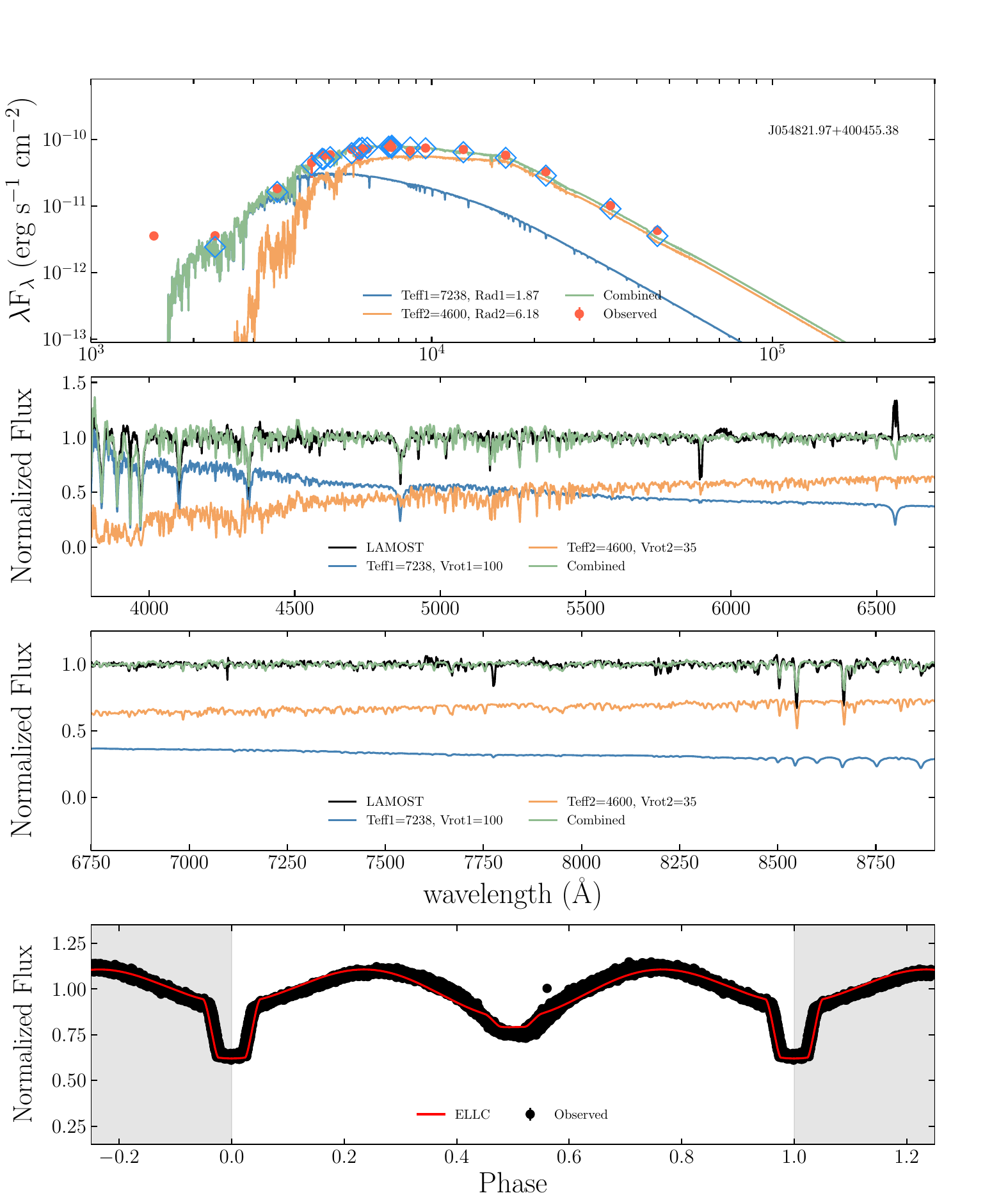}
\figsetgrpnote{ }
\figsetgrpend

\figsetgrpstart
\figsetgrpnum{3.10}
\figsetgrptitle{J0553
}
\figsetplot{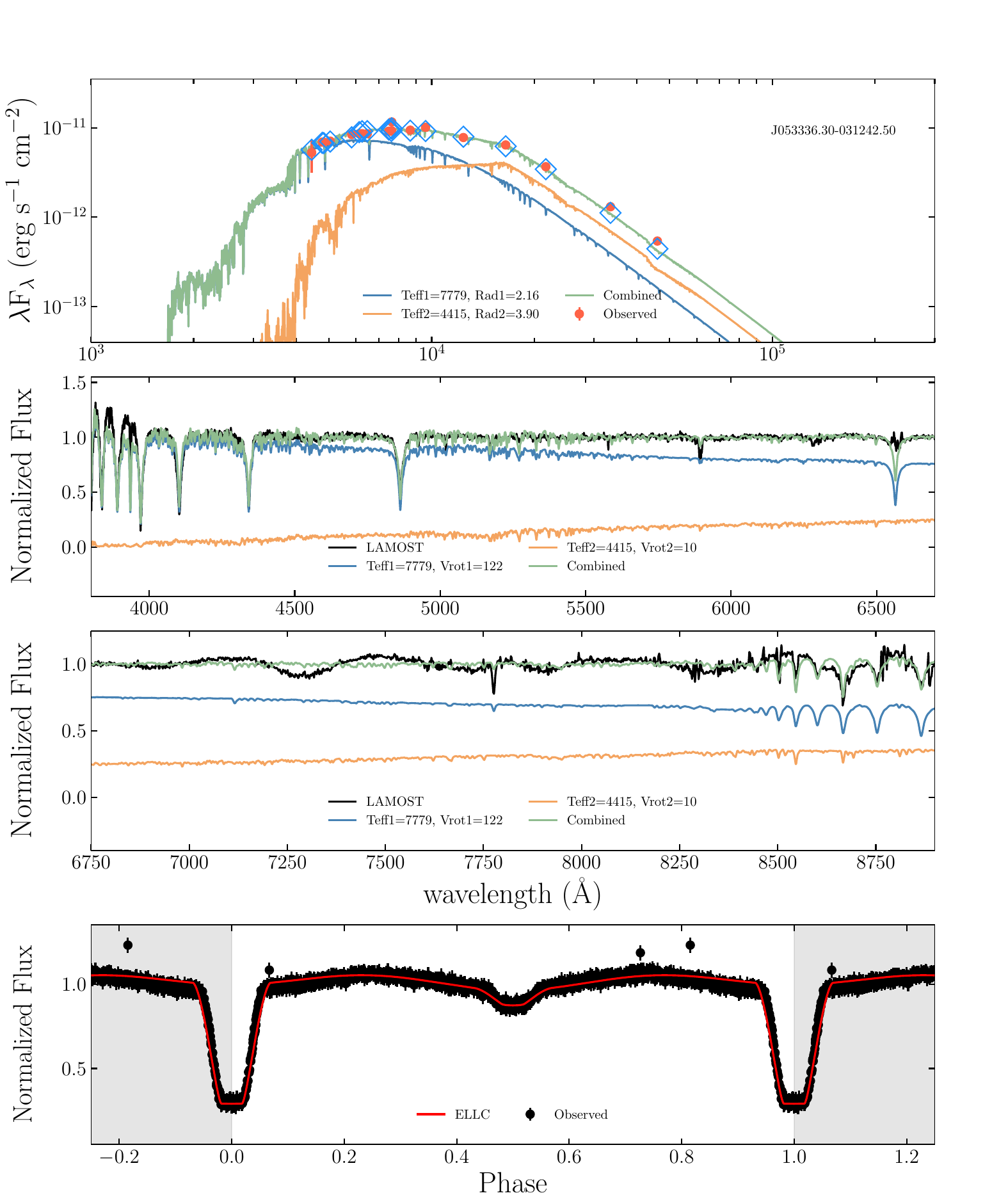}
\figsetgrpnote{ }
\figsetgrpend

\figsetgrpstart
\figsetgrpnum{3.11}
\figsetgrptitle{J0611
}
\figsetplot{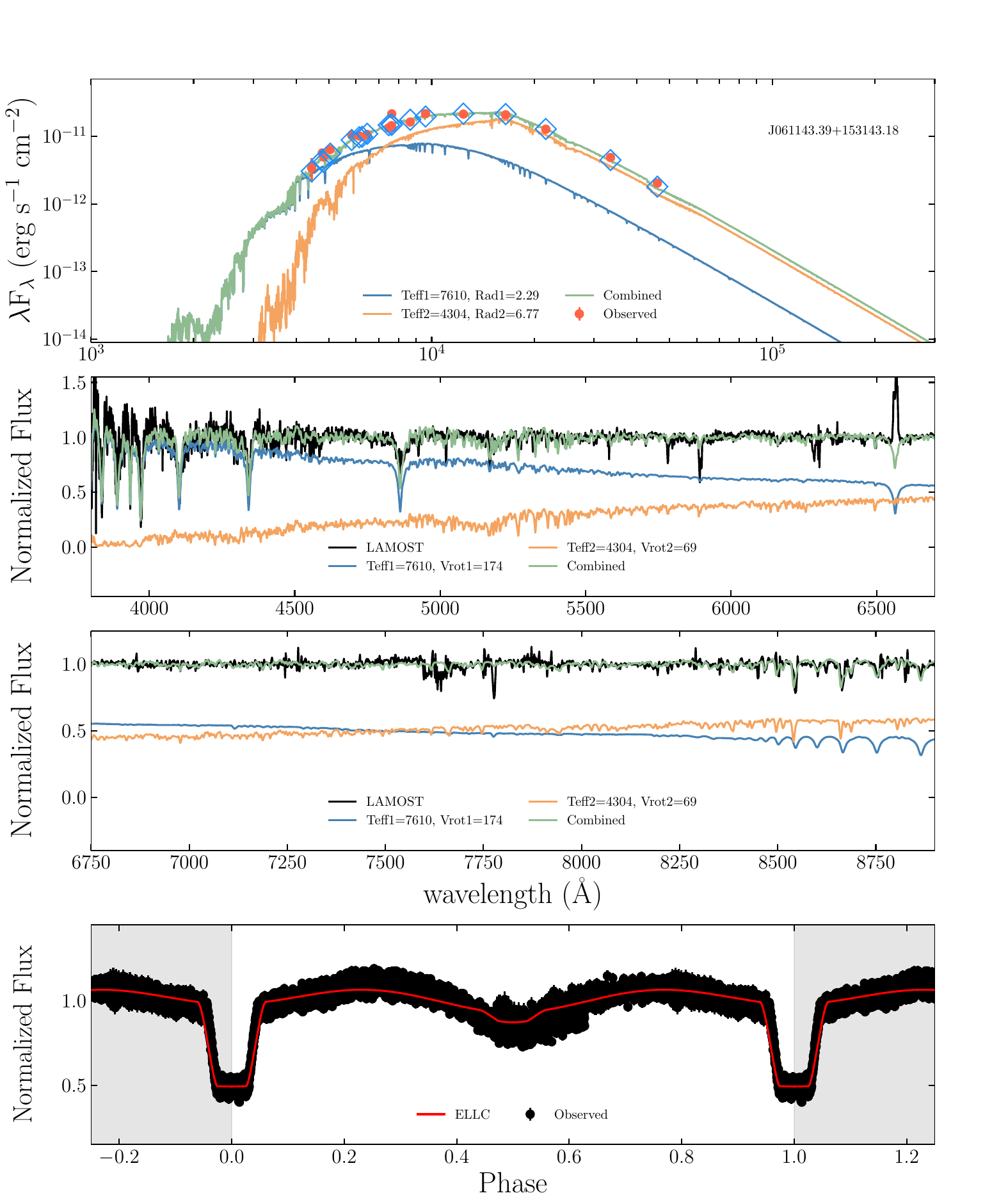}
\figsetgrpnote{ }
\figsetgrpend

\figsetgrpstart
\figsetgrpnum{3.12}
\figsetgrptitle{J0633
}
\figsetplot{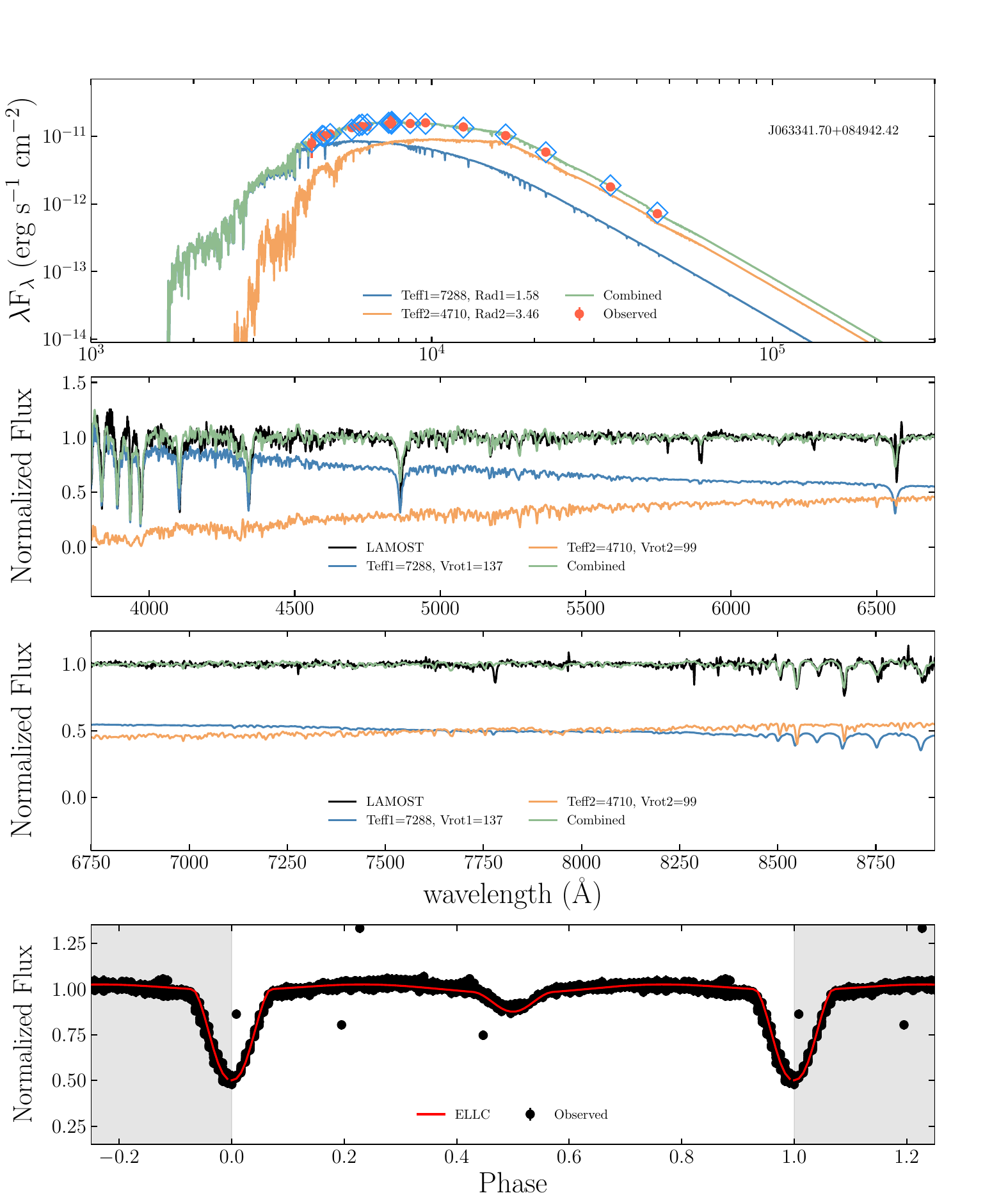}
\figsetgrpnote{ }
\figsetgrpend

\figsetgrpstart
\figsetgrpnum{3.13}
\figsetgrptitle{J0641
}
\figsetplot{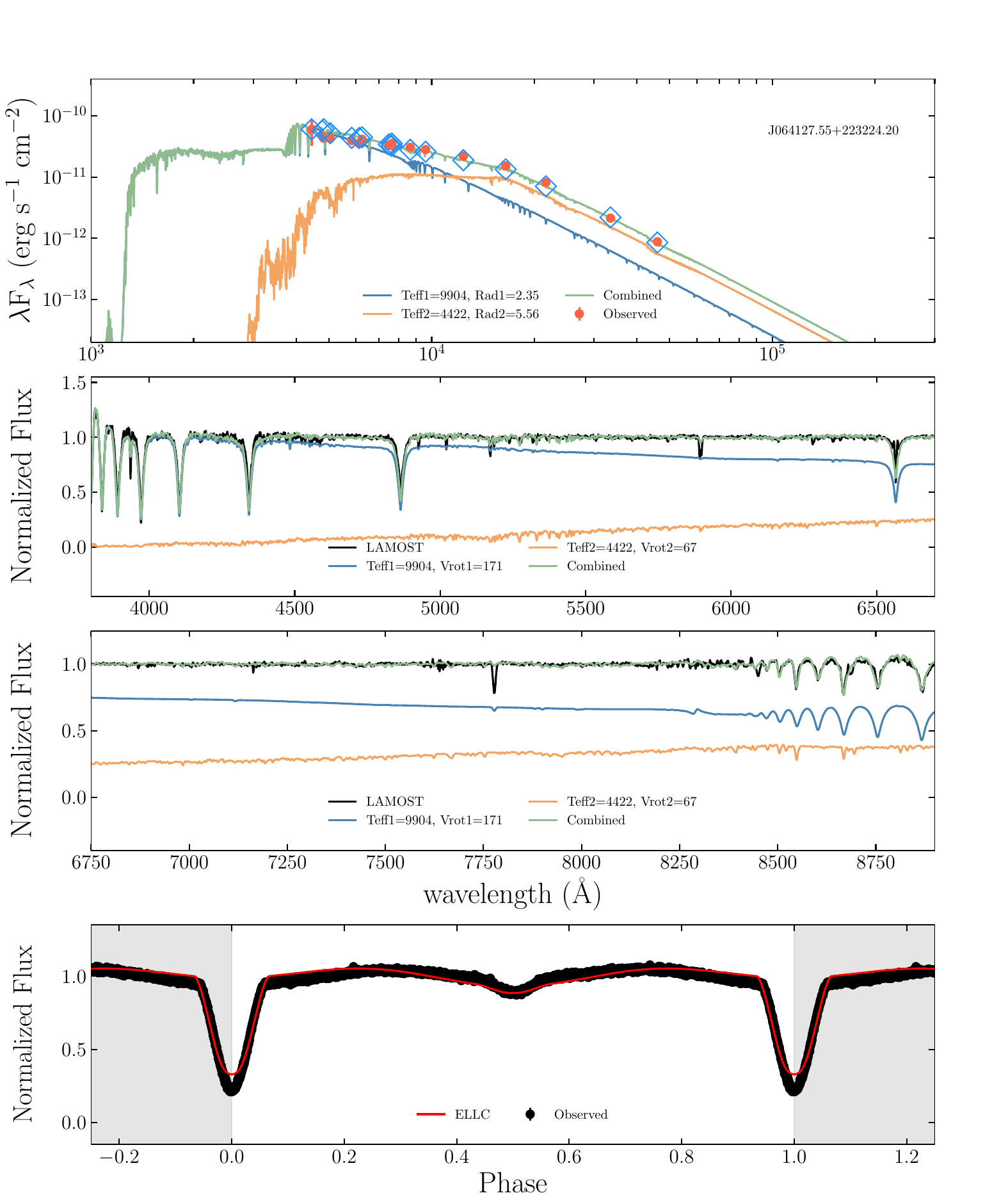}
\figsetgrpnote{ }
\figsetgrpend

\figsetgrpstart
\figsetgrpnum{3.14}
\figsetgrptitle{J0652
}
\figsetplot{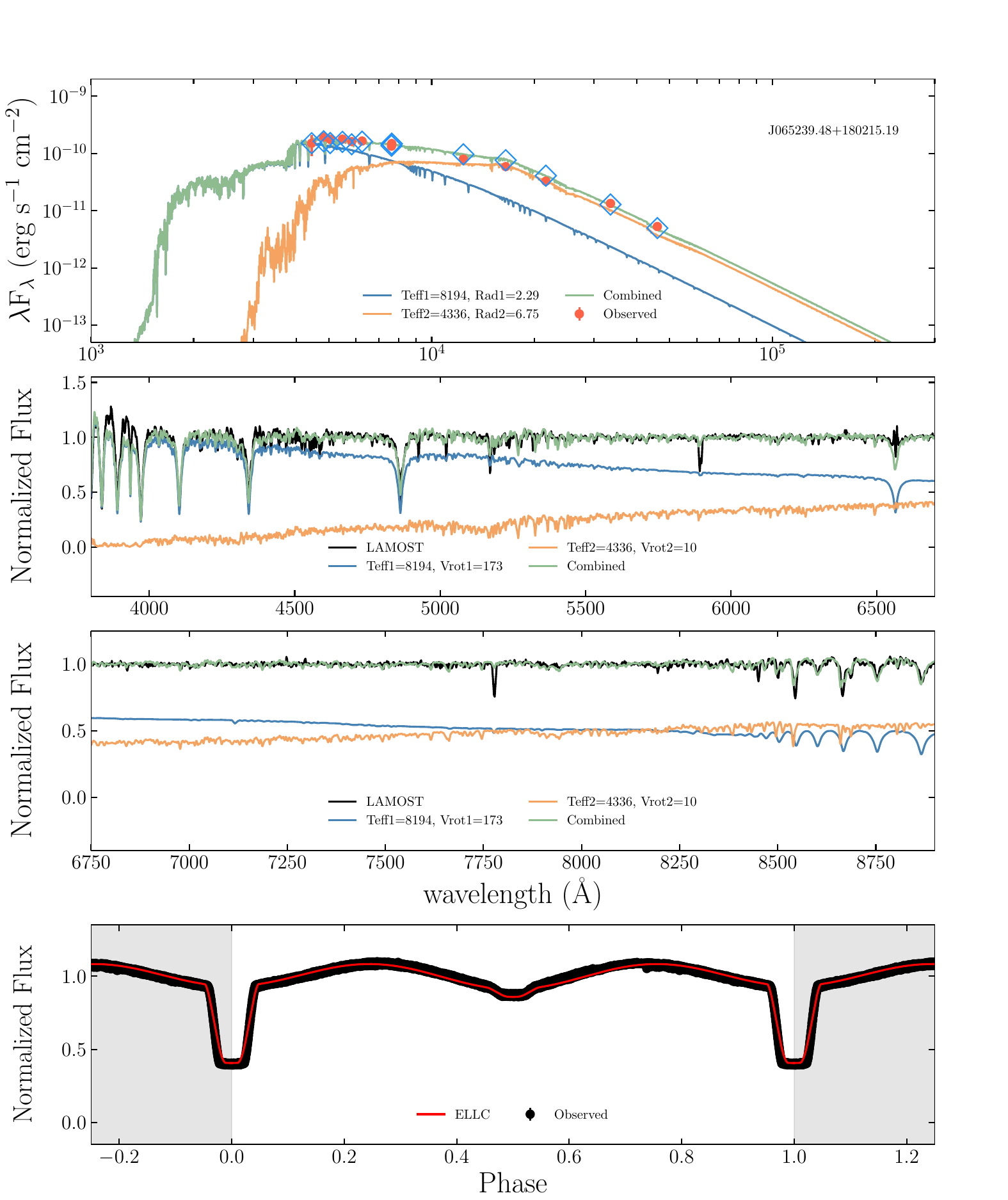}
\figsetgrpnote{ }
\figsetgrpend

\figsetgrpstart
\figsetgrpnum{3.15}
\figsetgrptitle{J0701
}
\figsetplot{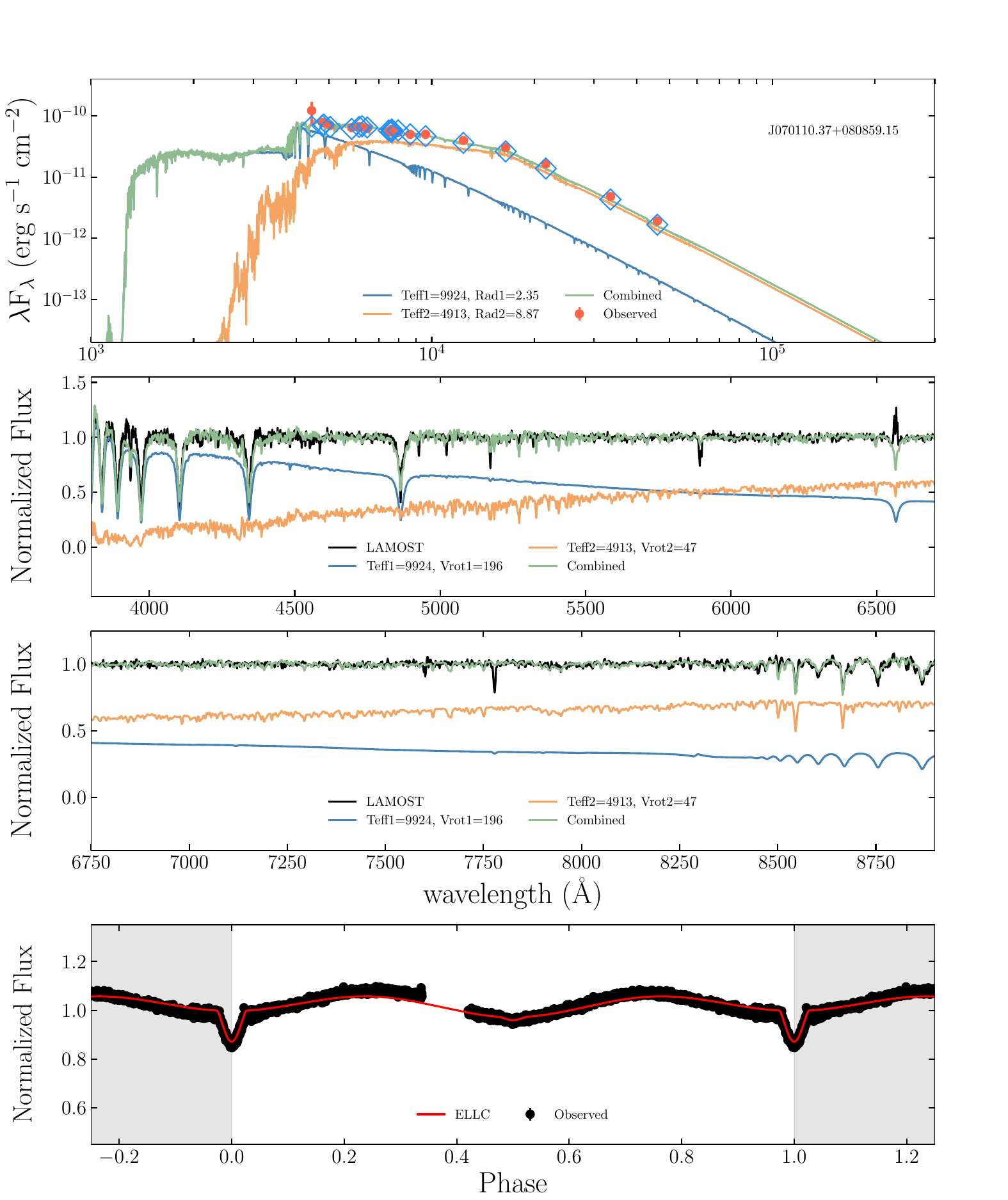}
\figsetgrpnote{ }
\figsetgrpend

\figsetgrpstart
\figsetgrpnum{3.16}
\figsetgrptitle{J0702
}
\figsetplot{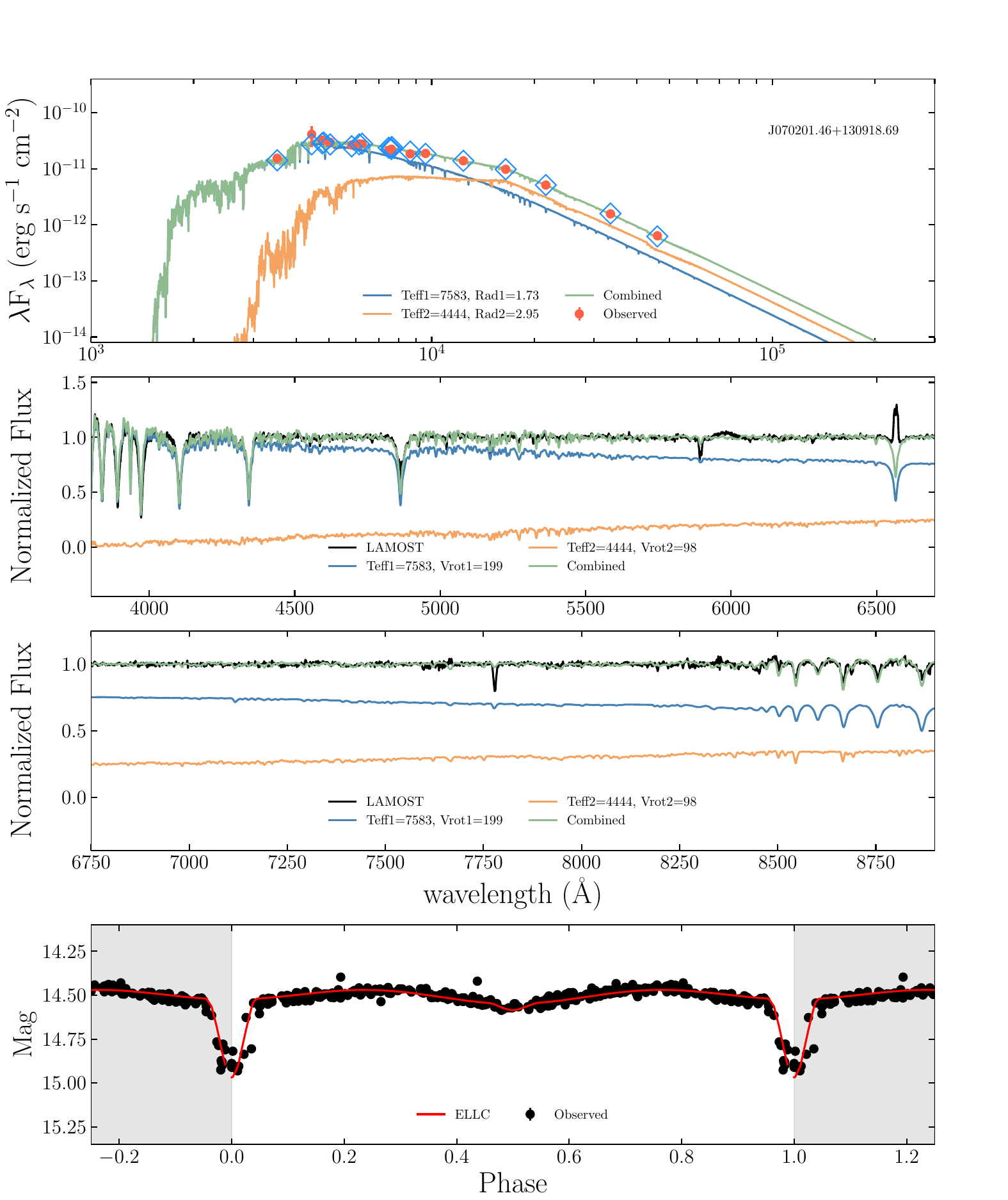}
\figsetgrpnote{ }
\figsetgrpend

\figsetgrpstart
\figsetgrpnum{3.17}
\figsetgrptitle{J0741.
}
\figsetplot{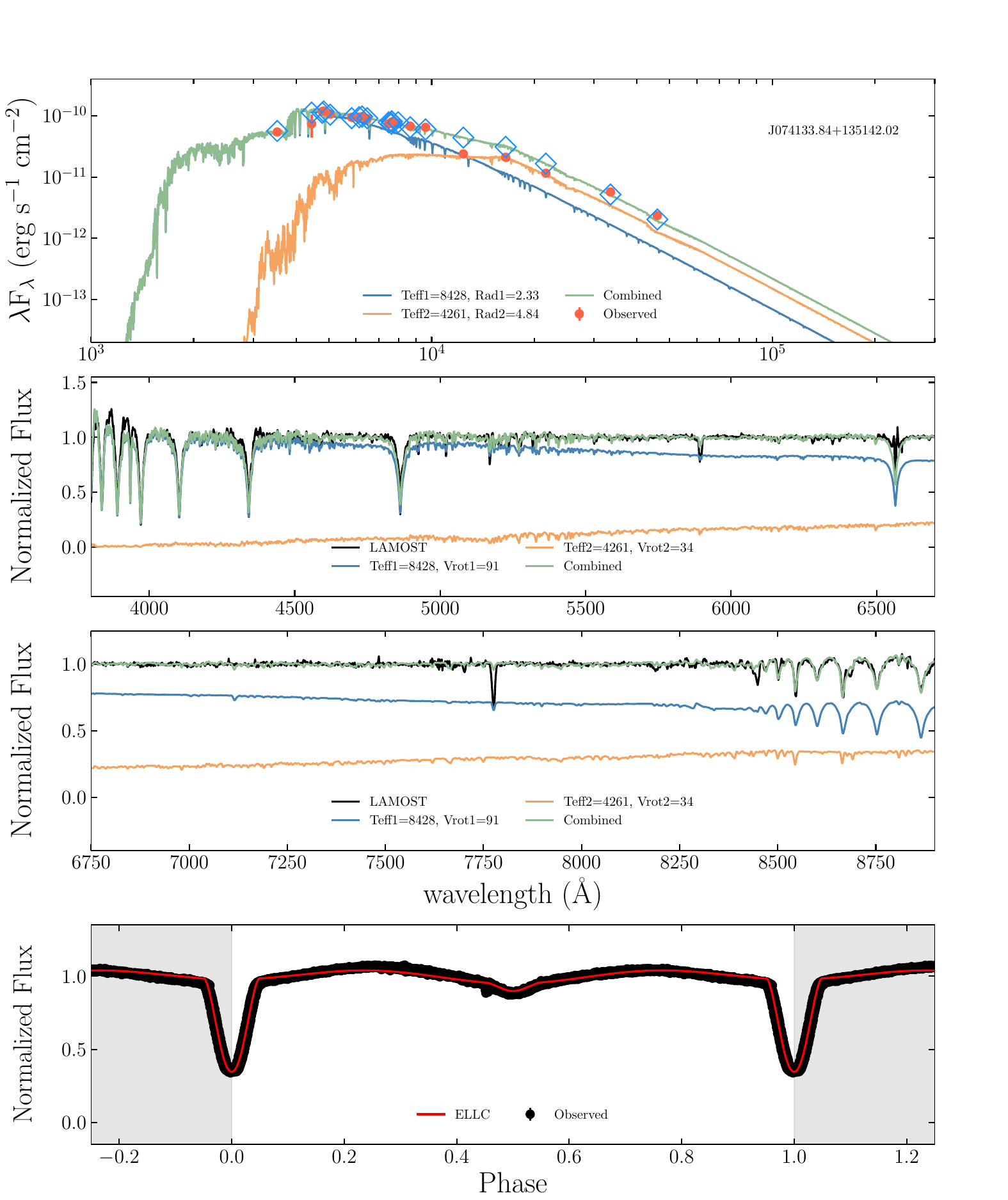}
\figsetgrpnote{ }
\figsetgrpend

\figsetgrpstart
\figsetgrpnum{3.18}
\figsetgrptitle{J1905
}
\figsetplot{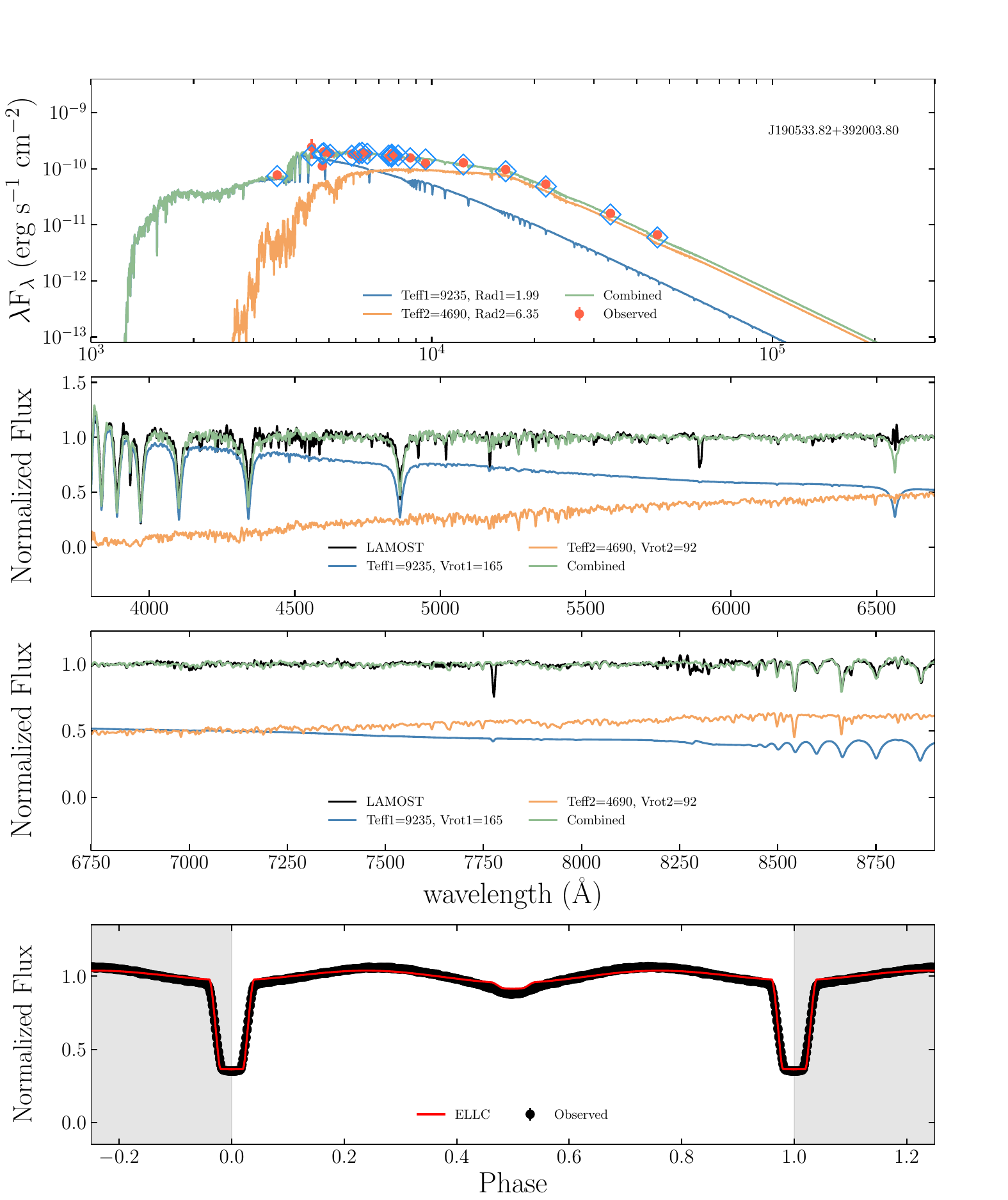}
\figsetgrpnote{ }
\figsetgrpend

\figsetgrpstart
\figsetgrpnum{3.19}
\figsetgrptitle{J1948
}
\figsetplot{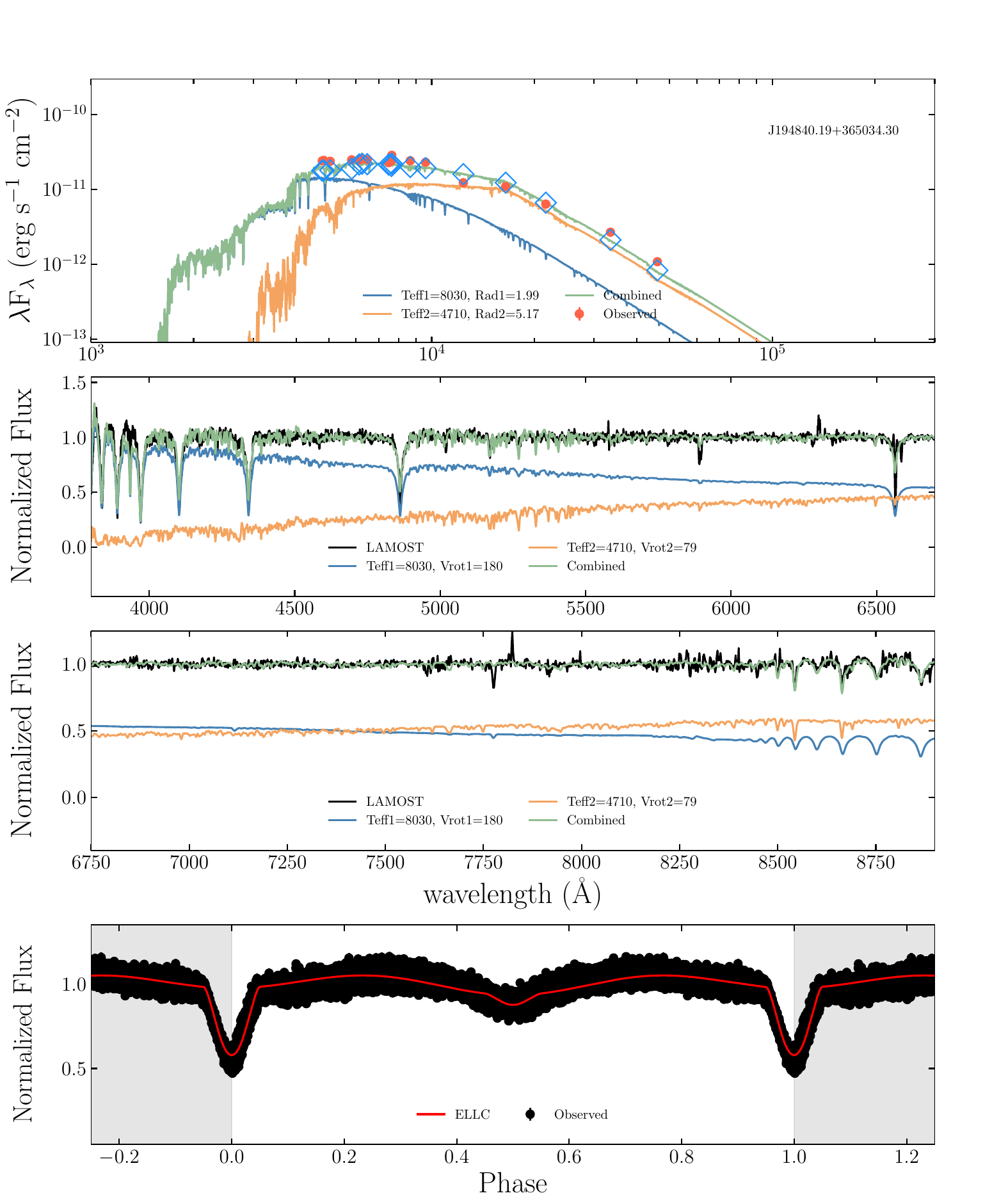}
\figsetgrpnote{ }
\figsetgrpend

\figsetgrpstart
\figsetgrpnum{3.20}
\figsetgrptitle{J2015
}
\figsetplot{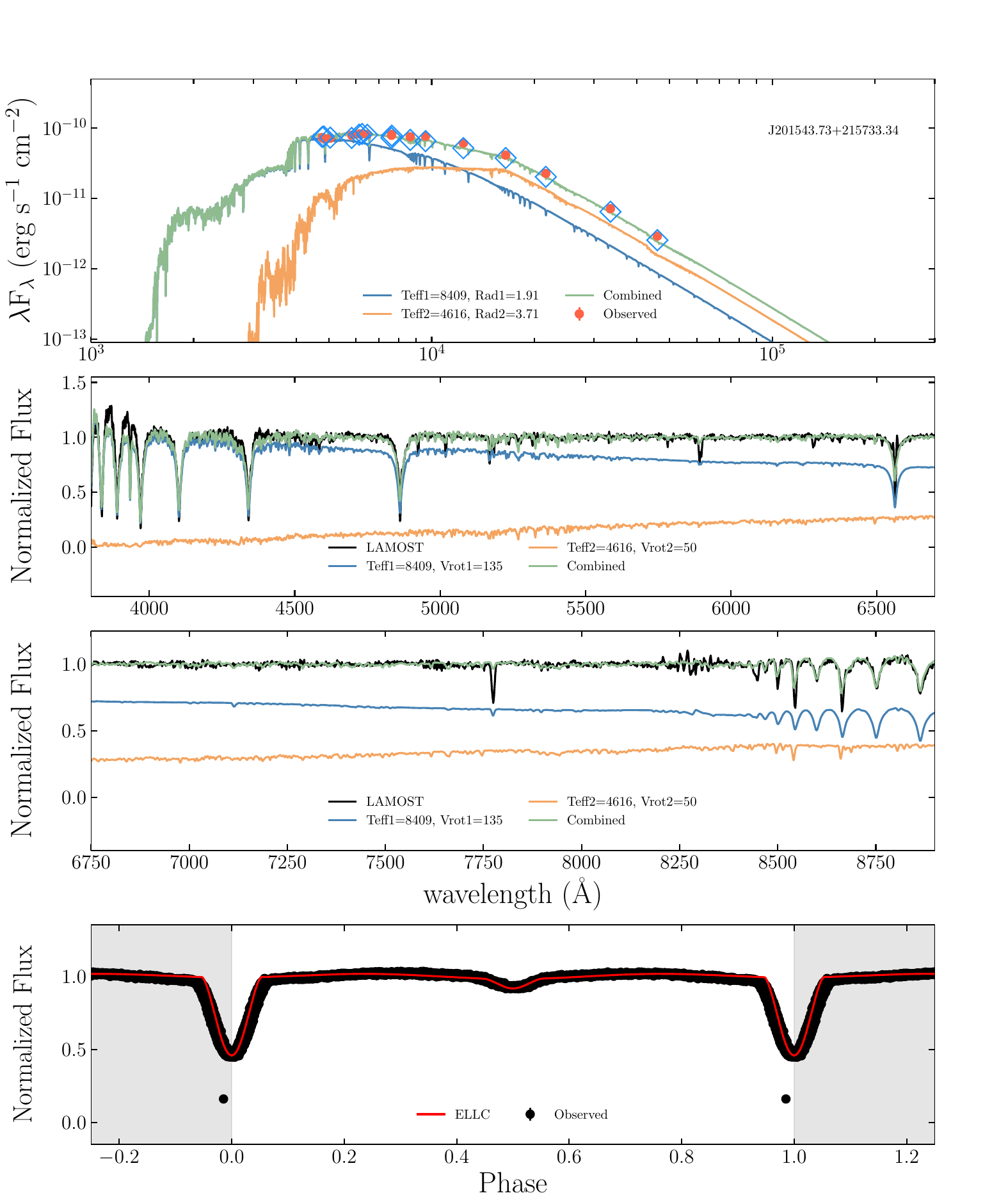}
\figsetgrpnote{ }
\figsetgrpend

\figsetgrpstart
\figsetgrpnum{3.21}
\figsetgrptitle{J2107
}
\figsetplot{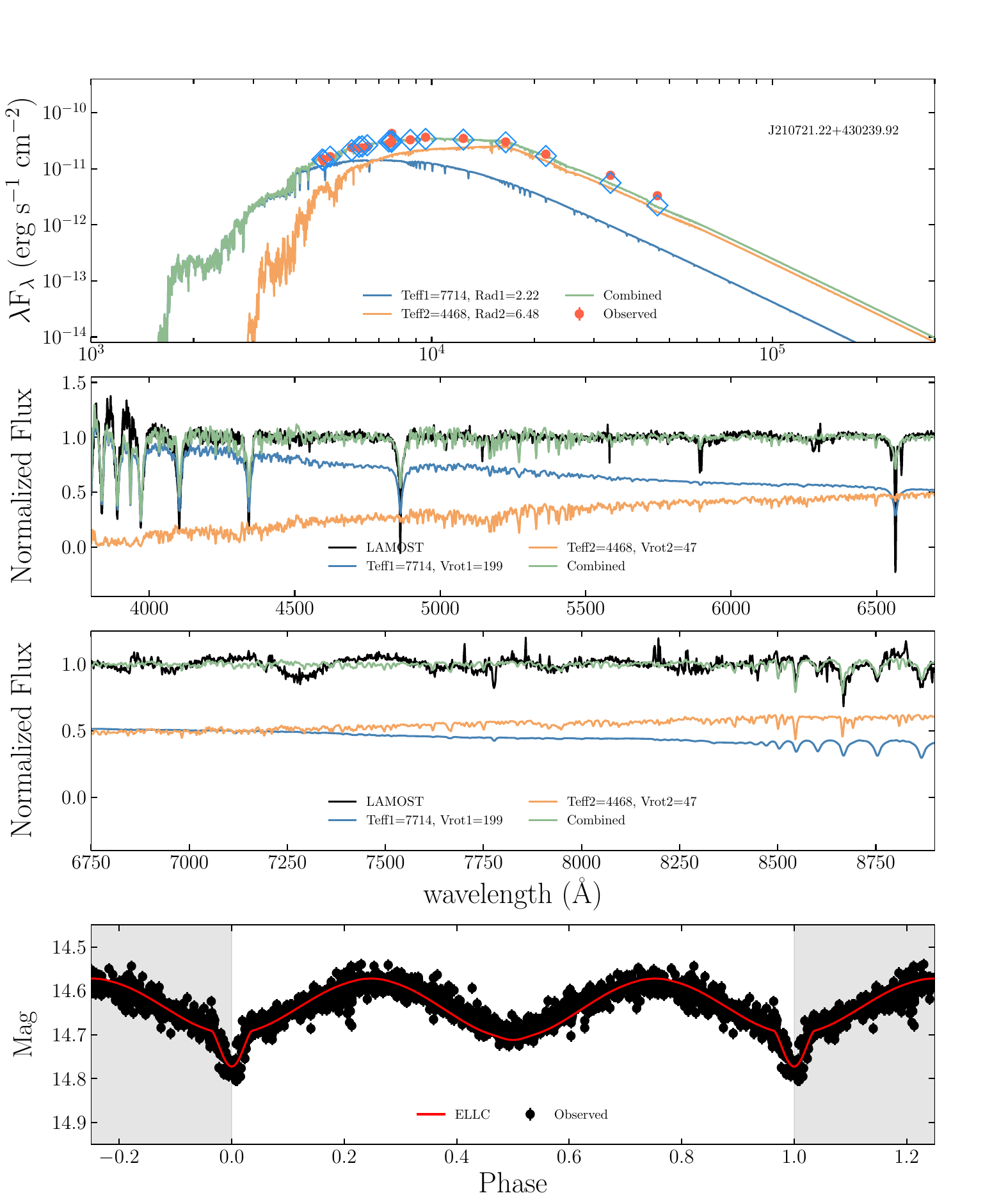}
\figsetgrpnote{ }
\figsetgrpend

\figsetgrpstart
\figsetgrpnum{3.22}
\figsetgrptitle{J2134
}
\figsetplot{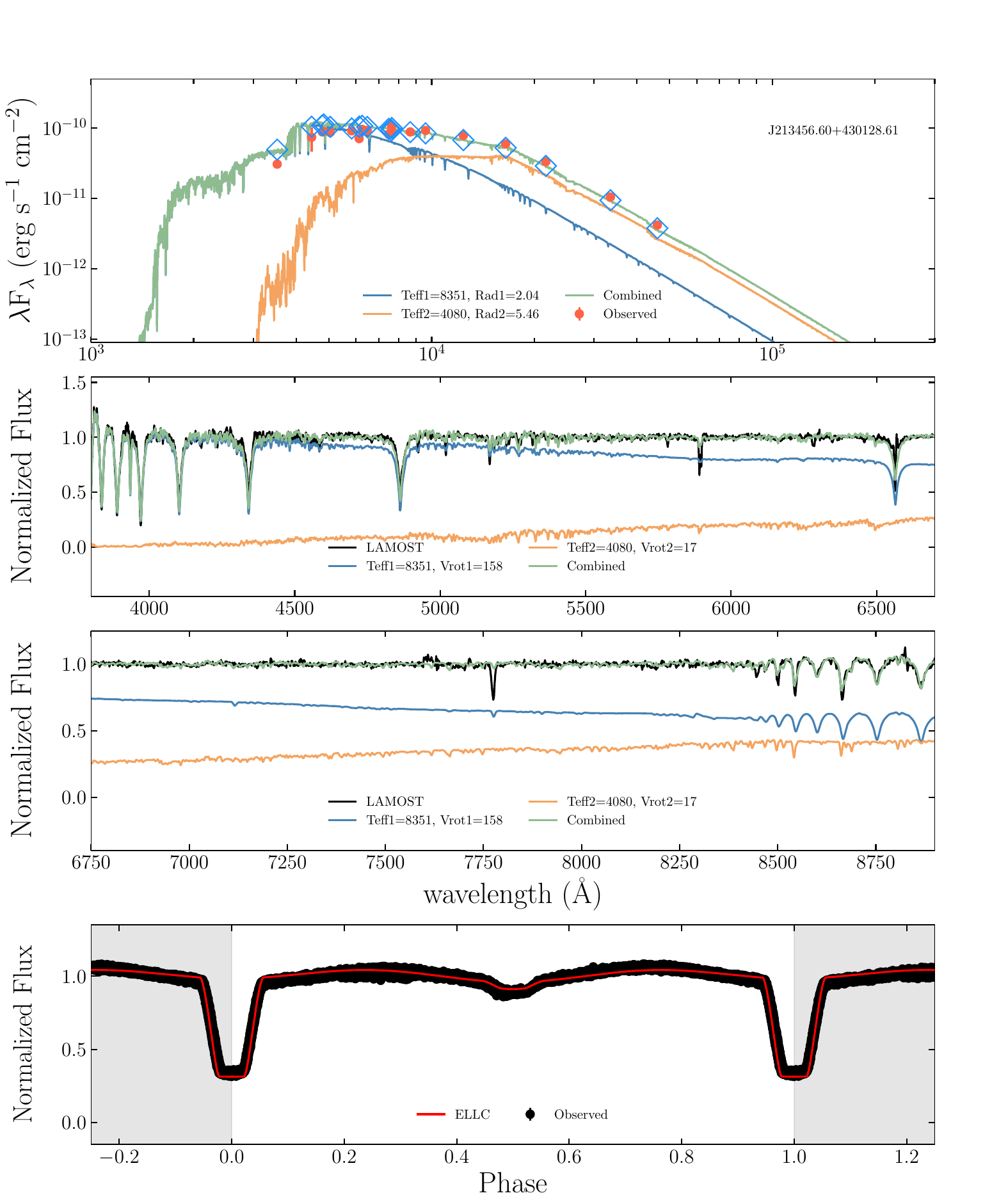}
\figsetgrpnote{ }
\figsetgrpend

\figsetgrpstart
\figsetgrpnum{3.23}
\figsetgrptitle{J2214
}
\figsetplot{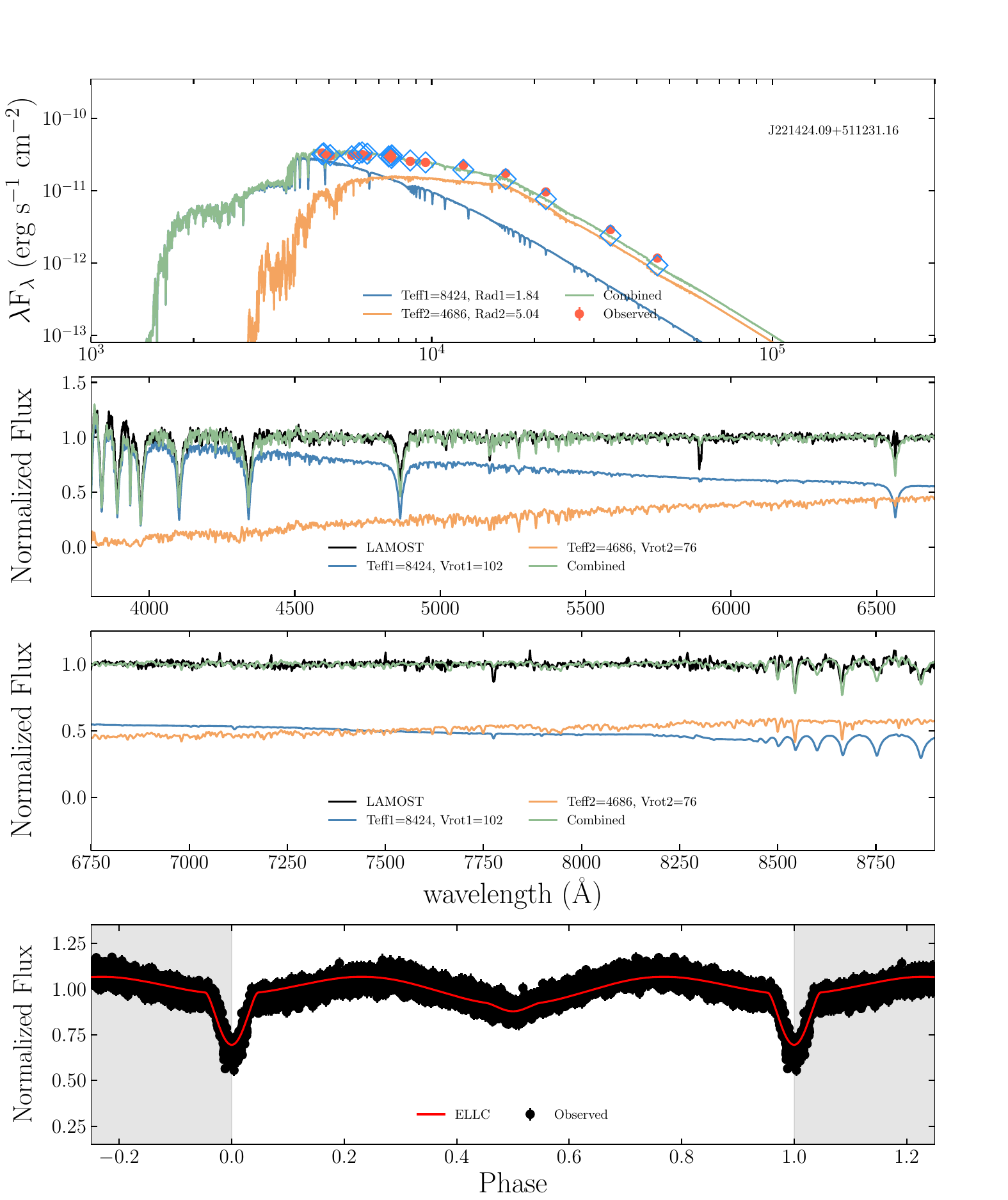}
\figsetgrpnote{ }
\figsetgrpend

\figsetend

\begin{figure*}
\plotone{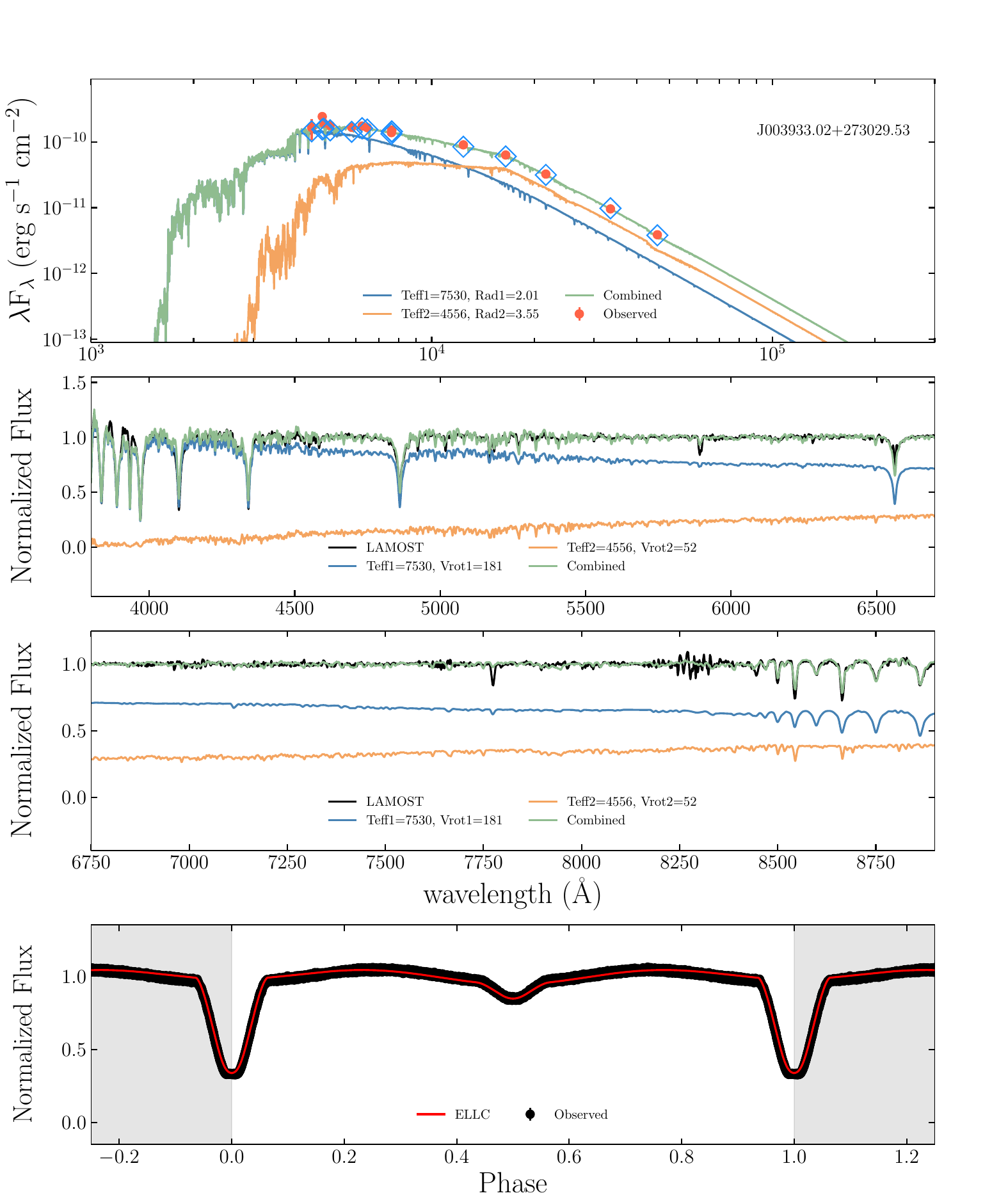}
\caption{Upper Panel: The SED fitting result. The red points illustrate the observed photometry, while the green line represents the combined SED model of J0039. The blue and orange lines depict the models for the accretor and donor, respectively. Middle Panels: The normalized LAMOST low-resolution spectrum and the best-fitting model. The green line shows the best spectral model calculated by the joint fit, while the blue and orange lines show the spectral models of the accretor and donor. Bottom Panel: The phase-folded light curve from the TESS band. The red line shows an \texttt{ellc} light curve model calculated by the joint fit. The complete figure set (23 images) is available in the online journal.}
\label{fig:fit_result}
\end{figure*}

To fit the spectral energy distributions (SEDs), we collected multi-band photometric data from various sky surveys, including GALEX \citep{martin2005galaxy,bianchi2011galex}, \Gaia \citep{vallenari2023gaia}, SDSS \citep{abazajian2009seventh}, APASS \citep{henden2014apass}, Pan-STARRS \citep{chambers2016pan}, TESS \citep{ricker2015transiting}, 2MASS \citep{skrutskie2006two}, and ALLWISE \citep{wright2010wide}, which cross a wavelength range of $0.2{-}10\,\mu\mathrm{m}$.

A single-star model is insufficient to fit the SEDs of the sample, particularly in the 
near-infrared range. The observed SEDs are better represented as a combination of two 
separate components: a hot accretor ($T_{\mathrm{eff}} \sim 8500\,\mathrm{K}$), which 
dominates the flux in the optical band, and a cool donor ($T_{\mathrm{eff}} \sim 4000\,\mathrm{K}$) with a large radius. Consequently, we utilized a composite model composed of two single-star templates. We generate the spectral templates using the 
following free parameters, where ``1'' refers to the accretor and ``2'' refers to the donor: 
effective temperature (\(T_\mathrm{eff,1}\), \(T_\mathrm{eff,2}\)), metallicity ([Fe/H]), 
surface gravity (\(\log{g_1}\), \(\log{g_2}\)), radius(\(R_1\), \(R_2\)), distance ($D$), 
and extinction (\(A_V\)), while also incorporating systematic errors. 
We assume that the metallicity of the two stars is the same.
Each single component of the SED model is interpolated by using a python package \texttt{stellarSpecModel}\footnote{\url{https://github.com/zhang-zhixiang/stellarSpecModel}} \citep{stellarSpecModel2025}
with given temperature, [Fe/H], and \(\log{g}\). The BT-Cond model grid \citep{Allard2012} is used when interpolating the stellar parameters.
We set a prior of the extinction (\(A_V\)) based on the 
result derived from the 3D dust map of \citet{green20193d}. An extinction correction is applied using the extinction model from \citet{cardelli1989relationship}, with a total-to-selective extinction ratio of \(R_V = 3.1\). 
The Python package \texttt{pyphot}\ \citep{2025pyphot} was utilized to convert spectral templates into synthetic photometry. This package provides the central wavelength, transmission curve, and zero point for a given filter. When calculating the radiation flux, we accounted for the contributions of both the accretor and the donor. The resulting synthetic photometry was then used to fit the observed data.

Stellar radii and distances are highly degenerate in the SEDs fitting process. To mitigate this, we use 
the zero-point-corrected parallaxes from {\it Gaia} DR3 \citep{vallenari2023gaia} as the prior of distance.
This approach provided a strong constraint on the total system luminosity, thereby improving the accuracy of the derived stellar radii. There are seven objects whose parallax measurements from \Gaia\ DR3 have uncertainties greater than 10\%, resulting in large uncertainties in radius estimates.

The temperature and extinction are strongly degenerate in the SEDs fitting process. Additionally, accurately constraining the temperature ratio between the binary components remains challenging and subject to significant uncertainties. To obtain more reliable fitting results, we performed a joint fitting of the SEDs, LAMOST spectra, and light curves in Section \ref{subsec:fit-result}.

Upon examining the SED fitting of J0741, we found that the three 2MASS data points exhibit an anomalous drop, see Fig. Set 3.17. This anomaly may have led to a substantial underestimation of the donor's radius, which in turn resulted in an underestimation of the donor's mass. To mitigate this potential influence, we removed the 2MASS data points from the subsequent SED analysis of J0741.

\subsection{Spectral fitting} \label{subsec:spec-fit}
Each of our sample sources has at least one stellar spectrum from LAMOST Low Resolution Survey (LRS; \(R \sim 1800\)) with a mean $\mathrm{SNR > 10}$. Among these, J0641, J0741, and J2214 were observed by LAMOST Medium Resolution Survey (MRS; $R \sim 7500$). Additionally, we also found APOGEE \citep{2019PASP..131e5001W} spectra for J1905. 
Before performing spectral fitting, we applied a continuum normalization procedure to the observational spectra. This step mitigates the influence of uncertainties in spectral slope calibration on the fitting results.

We used a two-component model to fit the observational spectra. Each component was also generated by interpolating the BT-Cond spectral grid using the \texttt{stellarSpecModel} package. And the interpolation parameters include effective temperature, metallicity, and surface gravity. The spectral model parameters were set to match those used in the SEDs fitting process.

Unlike the SED model, the spectral model requires additional parameters, such as radial velocities (RVs) and rotational broadening (\(v \sin i\)) of spectral lines. The RV and rotational velocities \(v \sin i\) of each component were treated as distinct free parameters in the spectral fitting. 
Star rotation causes different regions of the star surface to have varying radial velocities relative to the observer. The radiation emitted from these regions undergoes Doppler shifts, leading to the broadening of spectral lines. We used the \texttt{rotation\_filter} function from the \texttt{spectool} package\footnote{\url{https://github.com/zhang-zhixiang/spectool}} \citep{spectool2025} to broaden the spectrum to account for this effect.
The \texttt{rotation\_filter} function smooths the spectrum by applying a rotation kernel, which is defined as:
\begin{equation}
G(x) =
\begin{cases}
2(1 - \epsilon)(1 - x^2)^{1/2} + \frac{\pi \epsilon}{2}(1 - x^2) & \text{for } x \leq 1 \\
0 & \text{for } x > 1
\end{cases}
\end{equation}
where \( x = V / {v\sin i} \), $v\sin i$ is the projection of the star's maximum rotational velocity along the observer's line of sight, and \( \epsilon \) denotes the limb-darkening coefficient. The spectral model is described by the following equation:
\begin{equation}
Spec(\lambda) = \texttt{rotation\_filter}(\text{Model}( \lambda), v_{\text{rot}}, \text{limb}),
\end{equation}
where \( \text{Spec}(\lambda) \) represents the broadening template spectrum, while \( \text{Model}(\lambda) \) refers to the interpolated model spectrum. The parameter \( v_{\text{rot}} \) is the width of the rotation kernel, and \(\text{limb}\) denotes the limb darkening coefficient. Subsequently, we generated the combined model spectra as follows:
\begin{equation} 
F_{\text{cont}}(\lambda) = \left(\frac{R_{1}}{D}\right)^{2} \text{Spec}_{1}(\lambda) + \left(\frac{R_{2}}{D}\right)^{2} \text{Spec}_{2}(\lambda),
\end{equation}
where $\mathrm{Spec_{1}, Spec_{2}}$ refer to the synthetic spectra of both components. Then, we normalized our synthetic spectra and the observed spectra from LAMOST using
\begin{equation}
F_{\text{norm,model}}(\lambda) = \frac{F_{\text{cont}}(\lambda) }{P_{\text{model}}(\lambda)},
F_{\text{norm,obs}}(\lambda) = \frac{F_{\text{obs}}(\lambda) }{P_{\text{obs}}(\lambda)},
\end{equation}
where $P_{\text{model}}(\lambda)$, $P_{\text{obs}}(\lambda)$ represent the pseudo-continuum used to normalize our synthetic spectra and LAMOST spectra. 

\subsection{Light curve fitting} \label{subsec:lc-fit}
Each of our sample sources was observed by TESS or Kepler. However, five of these binaries have nearby stars that significantly affect the TESS measurements. To minimize the impact of the third light on these binaries, we utilized data from the ZTF survey. 

The difference in depth between the primary and secondary eclipses indicates a considerable temperature difference between the component stars, as shown in the phase-folded light curves in Figure~\ref{fig:lc}. Meanwhile, the binary star shows ellipsoidal modulation outside eclipse phases due to tidal distortions. 
We used the \texttt{ellc} package \citep{Maxted+16} to model the light curve of each binary system. The light curve model of \texttt{ellc} used in this paper includes six parameters: the primary radius (\(r_1\)), the secondary radius (\(r_2\)), the semi-major axis (\(a\)), the mass ratio (\(q = M_{2}/M_{1}\)), the inclination (\(i\)), and the surface brightness ratio (\(J\)), where \(J\) is the ratio of the secondary's surface brightness to the primary's in given band.
Algol-type binaries undergo rapid orbital circularization due to their Roche-lobe-filling configuration. In our sample, the equal time intervals between the two eclipses and the symmetric ellipsoidal variations strongly indicate that their orbits have also been fully circularized. Therefore, we fixed the eccentricity (\(e\)) at zero. The time of mid-eclipse (\(t_0\)) was determined from the luminosity minimum observed when no eclipse occurred. We obtain the limb darkening ($u$) coefficient and the gravity-darkening ($\tau$) coefficient in the TESS band or ZTF-r band from the tables of \citet{Claret2017Limb} and \citet{limbClaret2011}. 

The light curves of certain targets, such as J0547, exhibit asymmetric profiles and temporal changes, suggesting the presence of spot activity on the stellar surface. However, we did not make separate adjustments to the light curve fitting for these targets, such as adding spots into the light curve model, in order to ensure consistency in light curve analysis and to help prevent overfitting.

\subsection{Fitting and results} \label{subsec:fit-result}

We define the logarithmic likelihood function for each component (SED, spectral, and light curve fitting) with a generalized expression:
\begin{equation}
\label{eq:general_likelihood}
\ln \mathcal{L}_i = -\frac{1}{2} \sum \left[ \left( \frac{F_i - F_{\mathrm{m},i}}{\sigma_i} \right)^2 + \ln(2\pi \sigma_i^2) \right],
\end{equation}
where \(i \in \{\mathrm{SED}, \mathrm{SPEC}, \mathrm{LC}\}\) denotes the component type.  
For SED fitting: \(F_{\mathrm{SED}}\) is the observed photometric flux, \(F_{\mathrm{m,SED}}\) is the SED model flux, and the variance \(\sigma_{\mathrm{SED}}^2 = \sigma_{\mathrm{SED\_obs}}^2 + \sigma_{\mathrm{SED\_sys}}^2\) combines observational and systematic uncertainties. The summation is over all photometric bands. For spectral fitting: \(F_{\mathrm{SPEC}}\) is the normalized LAMOST spectral flux, \(F_{\mathrm{m,SPEC}}\) is the model spectral flux, and the total uncertainty in the spectral fitting is expressed as \(\sigma_{\mathrm{SPEC}}^2 = \sigma_{\mathrm{spec\_obs}}^2 + \sigma_{\mathrm{spec\_sys}}^2\). For light curve fitting: \(F_{\mathrm{LC}}\) and \(F_{\mathrm{m,LC}}\) are the observed and model fluxes of the light curve, with \(\sigma_{\mathrm{LC}}^2 = \sigma_{\mathrm{lc\_obs}}^2 + \sigma_{\mathrm{lc\_sys}}^2\) represent the uncertainties of the observed fluxes and the systematic uncertainties. The total logarithmic likelihood is the sum of contributions from all components:
\begin{equation}
\ln \mathcal{L}_\mathrm{total} = \ln \mathcal{L}_\mathrm{SED} + \ln \mathcal{L}_\mathrm{SPEC} + \ln \mathcal{L}_\mathrm{LC}.
\end{equation}
The posterior probability density was sampled by using the Markov Chain Monte Carlo Ensemble sampler \texttt{emcee} \citep{foreman2013emcee}. We ran 38 parallel chains, which is twice the number of free parameters, with each chain consisting of 10,000 steps. A comprehensive summary of all results is provided in Table~\ref{table:1}, and the fitting results are shown in Figure~\ref{fig:fit_result}.

The joint fit of SEDs, spectra and light curves provides robust constraints on the stellar and orbital parameters of the binary systems. The binaries in our sample consist of an A-type star and a low-temperature star with a large radius. In the optical band, the A-type star dominates the flux contribution, allowing the observed spectra to provide strong constraints on its temperature.

For eclipsing binaries with circular orbits, the ratio of the depths of the two eclipses depends on the temperature ratio of the two stars. Therefore, the temperature of the donor can be determined using the light curves in conjunction with the temperature of the A-type star derived from spectroscopy. With the temperatures of both components determined and a distance prior provided by \Gaia, the SEDs fitting effectively constrains the radii of the two stars.

The light curves not only help constrain the temperature ratio of the two stars but also provide constrain to ${\left(R_1 + R_2\right)}/{a}$, as well as the orbital inclination $i$. From the spectral and light curve characteristics, we infer that the donor is filling its Roche lobe. Under this condition, the radius of the donor can be used to estimate its mass (see Section \ref{subsec:mass_const}). Once the orbital semi-major axis (a), and the donor mass ($M_2$) are determined, the mass of the A-type star can be readily calculated. This comprehensive approach allows us to derive the stellar and orbital parameters of the binary system through the joint fitting process.
 
The fitting results indicate that infrared-dominant donors have temperatures ranging from 3800\,K to 5000\,K, and their radii are considerably larger than those of main sequence stars with similar temperatures, suggesting that the donors have become bloated.
Meanwhile, the accretor, surrounded by accreted mass, exhibits a higher temperature range of 7300\,K to 10000\,K, with most having masses greater than typical A- or F-type stars. 
Mass transfer causes stars in binary systems to develop characteristics distinct from those of main sequence stars.

\begin{deluxetable*}{cccrllcccc}
\tablecaption{Stellar parameters derived from joint fitting for the sources in our sample \label{table:1}}
\tablehead{
\colhead{ID} & \colhead{R.A.} &\colhead{Decl.} &\colhead{$P_{\rm{orb}}$} &\colhead{$T_1$} &\colhead{$T_2$} & \colhead{$R_1$} &\colhead{$R_2$} & \colhead{$M_1$} &\colhead{$M_2$}\\
\colhead{} & \colhead{} &\colhead{} &\colhead{(days)} &\colhead{(K)} &\colhead{(K)} & \colhead{$R_\odot$} &\colhead{$R_\odot$} & \colhead{$M_\odot$} &\colhead{$M_\odot$}
}
\decimalcolnumbers
\startdata
J0039&00:39:33.02&+27:30:29.5 & 4.4026 & $7530 \pm 29$ & $4557\pm 44$  & $2.01\pm 0.04$  & $3.55\pm  0.02$ & $2.01\pm 0.12$ & $0.31\pm 0.01$\\
J0239&02:39:35.15 &+57:35:08.4 & 3.6862 & $7617\pm 55 $ & $3765 \pm 81$ & $1.98\pm 0.06$  & $3.76\pm 0.10$ &$ 2.92\pm0.14$& $0.53 \pm 0.04$ \\
J0314&03:14:46.48&+46:45:45.0 & 3.3110 & $ 7506 \pm 45 $  & $4457 \pm 13 $ & $1.51 \pm 0.01$  & $2.53\pm 0.07$ &  $3.19\pm0.17$ & $0.20\pm 0.02$\\
J0340&03:40:49.14&+44:58:06.7 & 9.1925 & $9566\pm 92$ & $4694\pm 116$  & $1.59\pm 0.18$  & $6.99\pm 0.14$ & $ 1.26\pm 0.19$ & $0.55\pm 0.03$\\
J0423&04:23:32.57&+44:41:18.7 & 3.5207 & $7570\pm 29$ & $4155\pm 54$  & $1.87\pm 0.06$  & $4.02\pm 0.05$ & $3.54\pm 0.15$& $0.71\pm 0.03$\\
J0442&04:42:04.09&+27:40:21.2 & 16.8572 & $9704\pm 27$ & $4715\pm 23$  & $1.86\pm 0.07$  & $8.85\pm 0.07$ & $1.48\pm 0.13$& $0.33\pm 0.01$\\
J0533 & 05:33:36.30 & $-$03:12:42.5 & 4.1751 & $7779\pm 70$ & $4415\pm 51$  & $2.16\pm0.04$  & $3.90\pm 0.04$ & $2.52\pm0.11$&$0.46\pm0.02$\\
J0543&05:43:41.80&+13:21:35.9 & 3.2959 & $7432\pm 14 $ & $4009\pm 12$  & $1.58\pm 0.01$ & $2.89\pm 0.03$ &$2.71\pm0.14$& $0.30\pm 0.02$\\
J0547&05:47:09.83&+41:09:52.5 & 5.4123 & $7911\pm50 $  & $4802\pm 100 $  & $1.73\pm0.06$  & $4.19\pm0.11$ & $2.50\pm0.22$& $0.34\pm 0.03$\\
J0548&05:48:21.97&+40:04:55.4 & 11.7327 & $7238\pm 20$ & $4601\pm 18$  & $1.87\pm 0.04$  & $6.18\pm 0.05$  & $2.42\pm 0.10$ & $0.23\pm 0.01$ \\
J0611&06:11:43.39&+15:31:43.2 & 10.6612 & $7611\pm 27$ & $4304 \pm 29$ & $2.29\pm 0.07$  & $6.77\pm 0.07$  & $1.65\pm0.22$& $0.37\pm 0.01$\\
J0633&06:33:41.70&+08:49:42.4 & 2.5653 & $7288 \pm 49$ & $4710 \pm 42$  & $1.58\pm 0.05$  & $3.46\pm 0.08$  &$1.11\pm 0.38$& $0.85\pm 0.06$\\
J0641&06:41:27.54&+22:32:24.2 & 5.2576 & $9904 \pm 320$ & $4422\pm 175$ & $2.35\pm 0.39$  & $5.56\pm 0.49$  & $3.90\pm0.21$& $0.85\pm 0.24$\\
J0652&06:52:39.48&+18:02:15.2 & 12.3281 & $8194\pm 50$ & $4337\pm 60 $  & $2.29\pm 0.06$  & $6.75\pm 0.04$  & $3.50\pm0.05$& $0.27\pm 0.01$\\
J0701&07:01:10.37&+08:08:59.2 & 19.2054 & $9924\pm 186$ & $4913 \pm 79$  & $2.35\pm 0.05$ & $8.87\pm 0.56 $ & $1.28\pm0.06$&$0.26\pm0.05$\\
J0702&07:02:01.46&+13:09:18.7 & 4.0696 & $7584  \pm 18 $ & $4445 \pm 25 $ & $1.73\pm 0.03 $  & $2.95 \pm 0.01 $ & $2.86\pm0.25$&$0.21\pm0.01$\\
J0741&07:41:33.84&+13:51:42.0 & 6.6465 & $8428 \pm 26 $ & $4261 \pm 88 $  & $2.33\pm 0.05$  & $4.84\pm 0.05$  & $3.99\pm0.06$&$0.35\pm0.01$\\
J1905&19:05:33.82&+39:20:03.8 & 11.2578 & $9235\pm 65$ & $4690\pm 55 $  & $1.99\pm 0.05$  & $6.35\pm 0.10$  &$3.98\pm0.09$&$0.27\pm0.01$\\
J1948&19:48:40.19&+36:50:34.3 & 5.6512 & $8030\pm 56$  & $4711\pm 54 $  & $1.99\pm 0.11$  & $5.17\pm 0.13$& $2.98\pm0.13$&$0.59\pm0.05$\\
J2015&20:15:43.73&+21:57:33.3 & 3.5306 & $8409\pm 91$ & $4616\pm 125$  & $1.91\pm 0.52$  & $3.71\pm 0.45$  & $1.81\pm0.21$&$0.55\pm0.23$\\
J2107&21:07:21.22&+43:02:39.9 & 7.9467& $7714\pm 36$ & $4469\pm 42$  & $2.22\pm 0.06$  & $6.48\pm 0.08$ & $2.89\pm0.12$&$0.59\pm0.02$\\
J2134&21:34:56.60&+43:01:28.6 & 6.2744 & $8352\pm 84$  & $4080\pm 100$ & $2.04\pm 0.45$  & $5.46\pm0.34 $  & $3.85\pm0.13$&$0.56\pm0.11$\\
J2214&22:14:24.09&+51:12:31.2 & 6.8124 & $8424\pm 26$ & $4686\pm 69$  & $1.84\pm 0.07$  & $5.04\pm 0.26$  & $2.17\pm 0.18$&$0.37\pm0.06$\\ 
\enddata
\tablecomments{Column (1): ID; 
column (2): R.A,; 
Column (3): Decl.; 
column (4): orbital period from light curve; 
Column (5): effective temperature of the accretor measured by the fitting; 
column (6): effective temperature of the donor;
column (7): radius of the accretor;
column (8): radius of the donor;
column (9): mass of the accretor;
column (10): mass of the donor.
}
\end{deluxetable*}

\subsection{Mass constraints\label{subsec:mass_const}}
Given that the double-peaked H$\alpha$ emission lines (see Figure~\ref{fig:ha}) indicate ongoing mass transfer \citep{atwood2012modeling,2023A&A...670A..94R}, we infer that the donor stars have filled their Roche lobe. 
For a star filling its Roche lobe, the mean density (\(\bar{\rho}\)) is related to the orbital period ($P_\mathrm{orb}$) by the following equation \citep[see][]{Frank2002}:
\begin{equation}
\bar{\rho} \approx 110 P_{\text{hr}}^{-2} \ \text{g cm}^{-3}
\end{equation}
where \(P_{\text{hr}}\) is the orbital period in hours. The relationship between mean density and orbital period enables us to estimate the donor masses in these mass-transfer systems. The results are summarized in Table \ref{table:1}.
 
The findings reveal that approximately 90\% of the donor stars have masses ranging from 0.18 to 0.6 \( M_\odot \), which is too low to be explained by single-star evolution. It is evident that the donors in these binary systems have transferred the majority of their material to their companion stars. Based on the determined masses of the donors and the orbital semi-major axis (\(a\)) obtained from the joint fitting, we derived the masses of the accretors in this sample, which are larger than those of the donors. For further discussion on the masses and evolution of these binary systems, see Section \ref{subsec:discuss_mass}.  

\subsection{Individual component Color-Magnitude Diagram (CMD)\label{subsec:CMD}}

\begin{figure}
\includegraphics[width=0.45\textwidth]{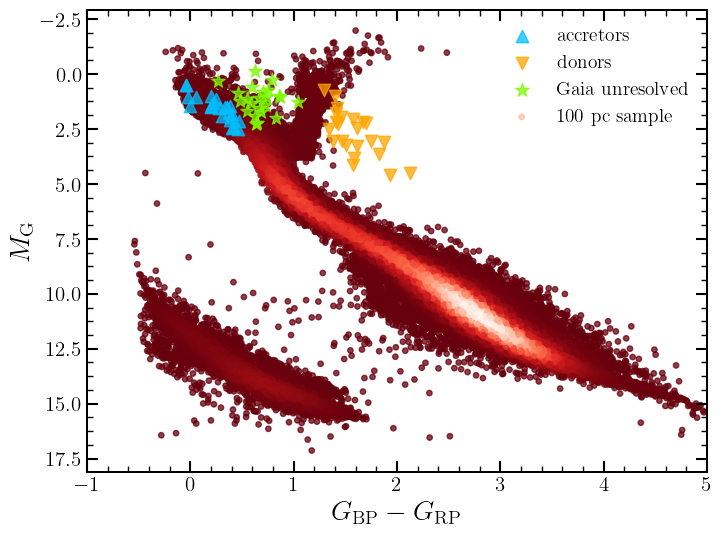}
\caption{Predicted positions of donors, accretors, and unresolved binaries on the Color-Magnitude Diagram (CMD). The positions of donors and accretors are inferred from the SEDs fitting. The blue triangles represent accretors, the orange inverted triangles represent donors, and the green circles indicate unresolved binaries obtained from \Gaia DR3 data. For comparison, we also plotted the CMD constructed from \Gaia DR3 observations that satisfy the filters described in Sect. 2.1 of \citet{2018A&A...616A..10G}, with a distance limit of 100 pc, as a background.
\label{fig:HR}}
\end{figure}

In Figure~\ref{fig:HR}, we separately plot the inferred Color-Magnitude Diagram (CMD) positions of accretors and donors. The temperatures and radii derived from the joint fitting suggest that the donors are located on or to the right of the the giant branch. In contrast, the accretors are situated close to the zero-age main sequence (ZAMS). 

Although the accretors appear to align along the main sequence in the CMD, it remains challenging to confirm whether they are currently in the main-sequence phase. As noted by \citet{Stevance2022}, the star that has gained a substantial amount of mass from its companion does not resemble a standard main-sequence star of the same mass in terms of brightness. Instead, it undergoes a phase of rejuvenation and ultimately appears significantly fainter.

We analyzed the masses and luminosities of these accretors and compared them with the mass-luminosity relation \citep{Eker2015,2018MNRAS.479.5491E}. The results reveal that the luminosities of most these objects are notably lower than those of main sequence stars with the same masses. This finding aligns with the predictions of \citet{Stevance2022}, further supporting the idea that the Algol systems studied in this work have not yet similar to the ordinary main sequence stars, although its CMD's position is align along the main sequence.
 
The donors in our sample are redder than main sequence stars and fainter than typical subgiant stars. They do not follow the standard evolutionary tracks for subgiants in the CMD. \citet{2017ApJ...842....1G} names these type objects as sub-subgiants (SSGs). They are formed through rapid stripping of a subgiant's envelope, or stellar collisions. Given that these donors have already lost the majority of their mass, it is reasonable that they deviate from the standard evolutionary paths. 

\citet{elbadry2022} reported a sample of binary systems with significant similarities to the objects studied in this work. However, in their sample, the donors appear to mostly align with the typical evolutionary tracks for giants. A possible explanation is that their constraints of the accretors and donors on the CMD, rely solely on SEDs fitting. In contrast, our calculations consider not only SEDs but also spectra and light curves. Different methods used to separate the binary components affect their positions.

\subsection{H$\alpha$ profiles and accretion disks}
\label{subsec:ha}

H$\alpha$ double-peaked emission features have been associated with the presence of disks in Algol-type binaries since they were first observed by \citet{wyse1934study}. Since then, one key piece of evidence for the presence of accretion disks in Algols is the observation of double peak emission lines during eclipses \citep{carpenter1930u,wyse1934study,atwood2012modeling,2020A&A...644A.121L}. All sources in our sample exhibit double-peaked H$\alpha$ emissions in the original LAMOST spectra, as shown in Figure~\ref{fig:ha}. We determined the stellar and orbital parameters by jointly fitting the SEDs, spectra, and light curves, as described in Section \ref{subsec:fit-result}. The residuals, calculated by subtracting a composite theoretical photospheric spectrum
from the observed data, where the composite theoretical photospheric spectrum is the spectral fitting results in Section \ref{subsec:fit-result},
reveal both single- and double-peaked H$\alpha$ emissions, along with other features. These H$\alpha$ profiles that have subtracted the continuums are shown in Figure~\ref{fig:ha}, represented by the gray curve. 

\begin{figure*}
\plotone{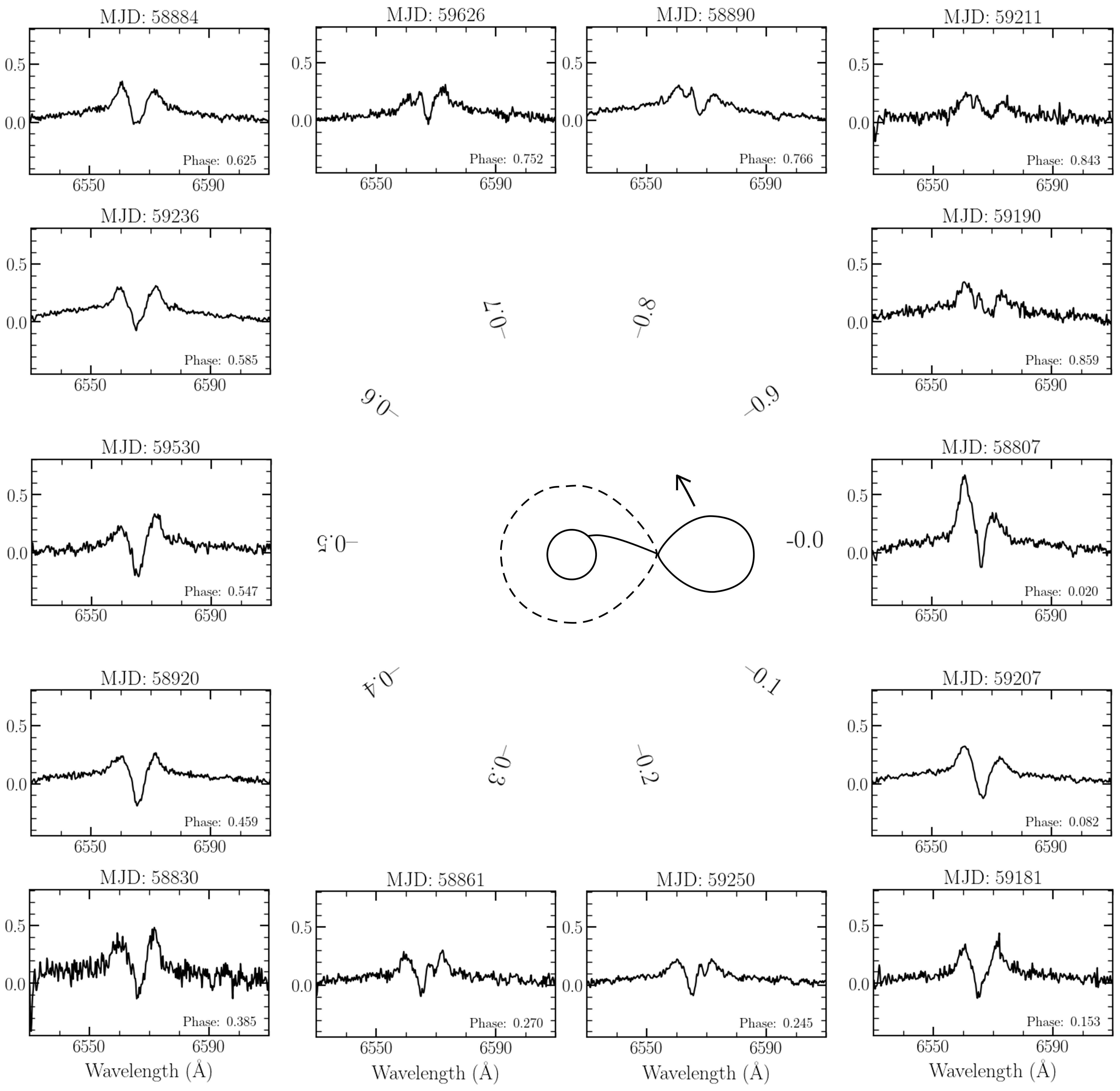}
\caption{Observed profiles of J0641's H$\alpha$ emission from LAMOST MRS. The H$\alpha$ emission line shows variation with the orbital phase.
\label{fig:0641}}
\end{figure*}

The analysis by \citet{richards1999morphologies} of the H$\alpha$ difference profiles reveals that accretion structures in Algol binaries exhibit four primary morphological types. The dominant morphologies are double-peaked emission systems and single-peaked emission systems. Double-peaked emission systems feature a transient or classical accretion disk, characterized by a disk-like distribution of the accreted material around the accretors. Single-peaked emission systems exhibit a gas stream-like distribution. The remaining types include alternating single- and double-peaked emission systems and weak spectrum systems.

Most of our sample sources exhibit H$\alpha$ double-peaked emission, which is commonly observed in a long period ($P_\mathrm{orb} > 6 \, \text{d}$) Algol systems \citep[e.g., see][]{budaj2005study, atwood2012modeling, zhou2018occultations}. In these systems, the viscous processes in the disk cause the gas to spread within the orbital plane, resulting in the formation of a stable, classical accretion disk. Accretion disks in longer-period systems produce emission lines that are much stronger and more permanent than those with shorter periods. In our sample, there are 11 long-period systems, all of which exhibit H$\alpha$ double-peak emission, as shown in Figure~\ref{fig:ha}.

Moreover, not only do long-period Algol systems in our sample exhibit H$\alpha$ double-peaked emission, but three intermediate Algol systems (J0547, J0641, J1948) with periods between 4.5 and 6 days also display this characteristic. However, their formation mechanisms differ significantly; the emission lines in intermediate-period systems originate from transient disks rather than classical disks. The line profiles from a transient disk appear broadened by supersonic turbulence, while the dominant broadening mechanism in a classical disk is the near-Keplerian motion of the gas within the disk. Furthermore, the gas in a transient disk is highly variable and may vanish in as little as one orbit, whereas the emission is a permanent feature of a classical disk \citep{1989Spectroscopic}. A prominent example is the source J0547, where evidence of this accretion phenomenon is clearly visible in its spectra. During its observations, H$\alpha$ was prominently seen in emission, with an equivalent width of $23.10 \pm 0.05 \, \text{Å}$ measured from the LAMOST spectra at phase \(\phi = 0.41\). At the same phase, the H$\beta$ and $\text{Ca II triplet (CaT)}$ lines display inverse P-Cygni profiles. Inverse P-Cygni profiles are spectral features where the absorption component appears on the red side of the emission lines, which may indicate the presence of inflowing gas from regions near the accretion disk \citep{2024ApJ...977L..28H,2013MNRAS.436..511M}. However, at phase \(\phi = 0.56\), the H$\beta$ and Ca II triplet (CaT) lines disappeared. This provides further evidence for the existence of transient disks. Due to the limited number of observations, there is insufficient evidence to discuss J0547 as an alternating single- and double-peaked emission system. J0641 was observed with LAMOST over 23 epochs from December 19, 2019, to February 15, 2022, at medium resolution and over 19 epochs from January 13, 2018, to March 5, 2022, in medium resolution survey. These numerous observations almost entirely cover the phase, as shown in Figure~\ref{fig:0641}. For J0641, the blue-shifted and red-shifted emission peaks show opposite trends at phases 0.0 and 0.5. These variations indicate the presence of an accretion disk in this Algol-type binary system. At phase 0.0, the donor star eclipses the disk around the accretor. Conversely, at phase 0.5, the disk, rotating with the accretor, is closest to our line of sight. This directly causes systematic changes in the emission line profiles.

The most common type in short-period Algols with $P_\mathrm{orb} < 4.5 \, \text{days}$ is a predominantly single-peaked emission feature in the H$\alpha$ difference profiles \citep{richards1999morphologies}. Studies using 2D and 3D Doppler tomograms \citep{richards2004doppler, richards2012new} explain that the H$\alpha$ emission lines in various short-period Algol binaries may originate from structures such as the gas stream, the chromosphere and other magnetic structures on the donor star, an accretion annulus, and an absorption zone. The short-period ($P_\mathrm{orb} < 4.5 \, \text{days}$) Algol binaries in our sample exhibit not only single-peaked H$\alpha$ emission, but some of them also show double-peaked emissions. A less common morphological type in the group of short-period Algols was a widely double-peaked H$\alpha$ emission, which suggests a disk-like distribution where the gas is in a transient or classical accretion disk \citep{2004A}. However, in our sample, this type of emission is more prevalent, which is likely due to a selection effect, as the distinct emission structure would be more easily observed. 

Nonetheless, due to the limited number of observations and the influence of phase effects, our current analysis cannot give a definitive conclusion regarding the morphological type of H$\alpha$ emission, except for J0641, which has spectral observations across the entire phase.

\subsection{Are most of the objects oscillating Eclipsing Algol systems?\label{subsec:pulsation}}

\begin{figure}
\includegraphics[width=0.465\textwidth]{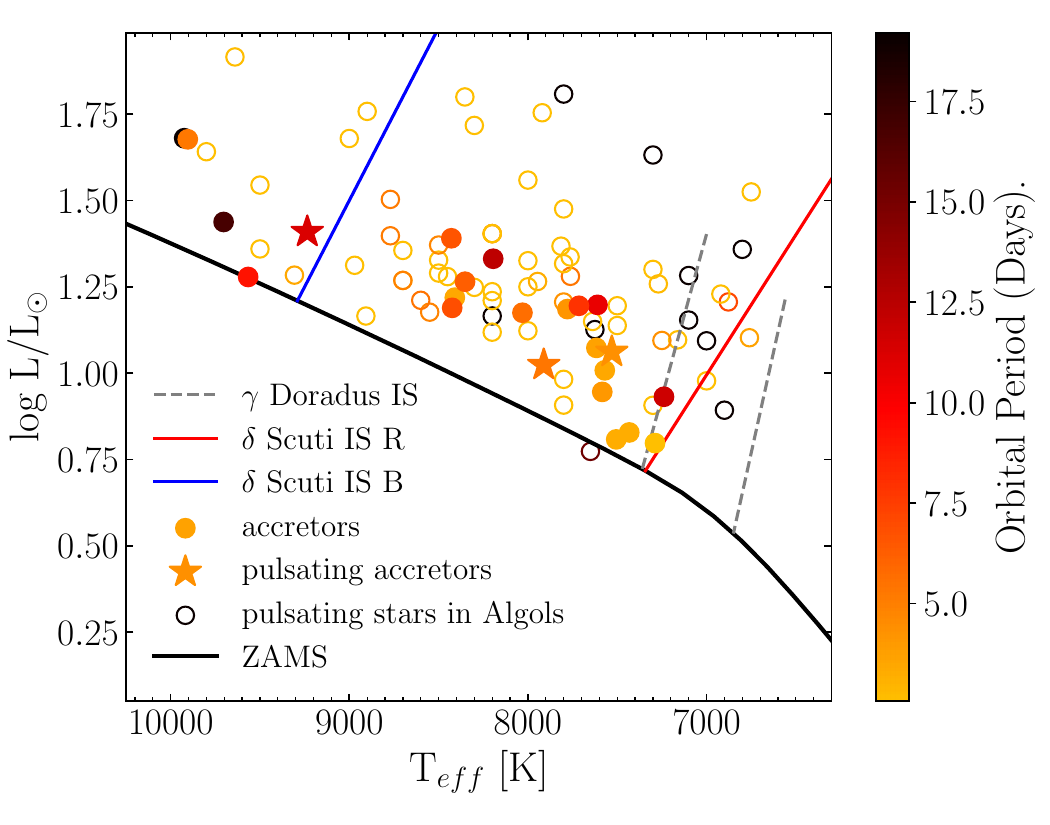}
\caption{A HR diagram showing the positions of the accretors in our sample. The red and blue lines represent the edges of the instability strip for the $\delta$ Scuti stars, while the black line indicates the theoretical Zero-Age Main Sequence (ZAMS). These data are based on the work of \citet{murphy2019gaia}. The gray dashed line represents the edge of the \(\gamma\) Doradus instability strip from \citet{2003ApJ}. The filled circles represent sources in the sample where pulsation features have not yet been detected. The filled stars refer to sources from our sample that have already exhibited pulsations in their light curves. The open circles indicate the pulsating stars in Algols from \citet{2017Liakos}. The color bar displays the orbital periods of all binary systems.}
\label{fig:HRD}
\end{figure}

\begin{figure*}
\centering
\includegraphics[width=0.95\textwidth]{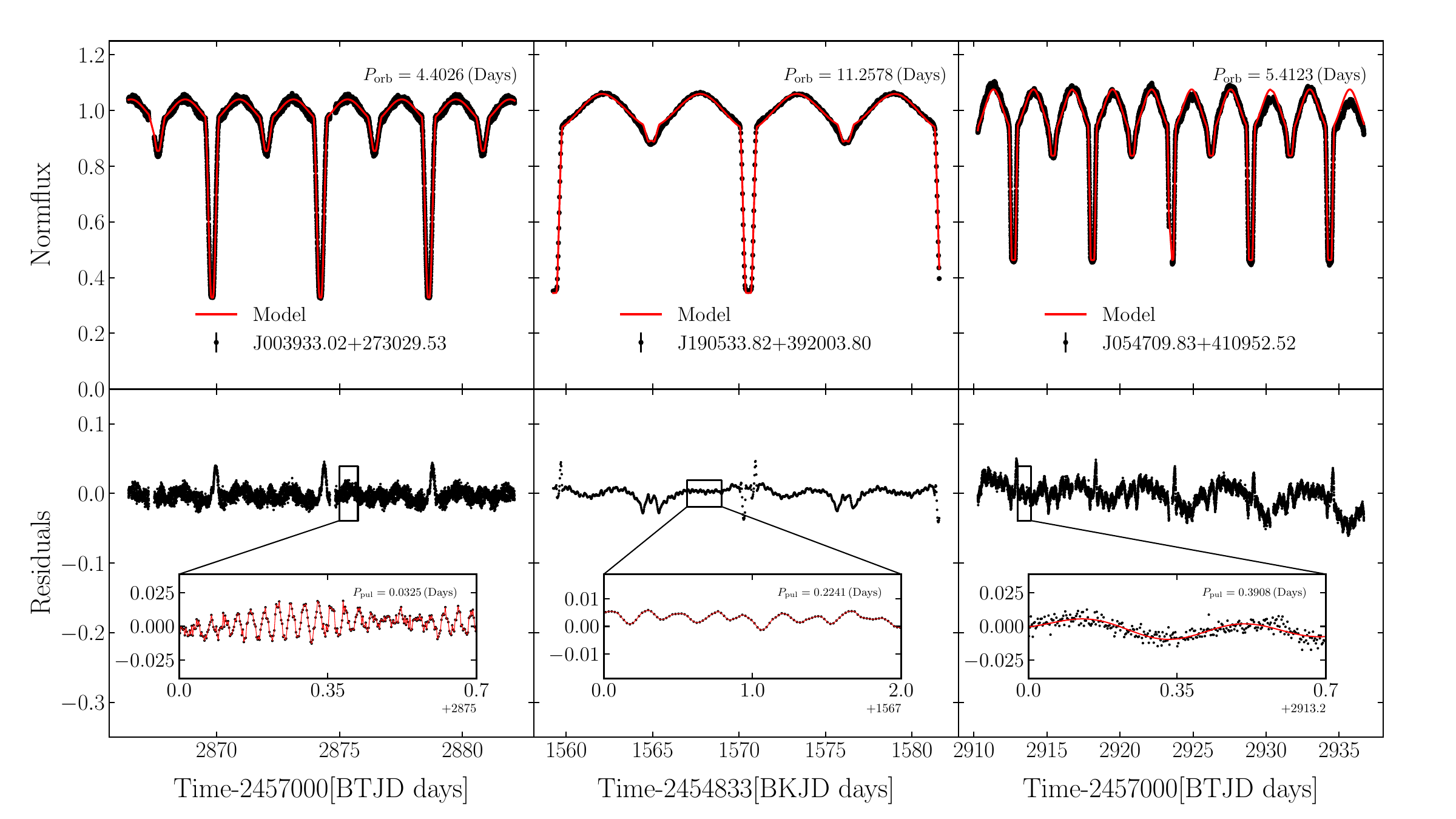}
\caption{The top panels display the TESS light curves of J0039 and J0547, along with the Kepler light curve of J1905. The red lines represent their \texttt{ellc} light curve models. The residuals from the fits are shown in the bottom panels, including the zoomed-in residuals of the light curves over a range of 0.7 to 2.0 days.
\label{fig:pulsation}}
\end{figure*}

An increasing number of \(\delta\) Scuti stars, frequently observed in Algol-type binary systems, have been found to exhibit short pulsation periods and higher amplitudes \citep{pigulski2007pulsating}. 
Therefore, we examined whether the accretors in our sample are located at the $\delta$ Scuti instability strip as proposed by \citet{murphy2019gaia}, as illustrated in Figure~\ref{fig:HRD}.
We found that 18 of our accretors lie within the $\delta$ Scuti instability strip. Additionally, we investigated whether any of the accretors are located in the $\gamma$ Doradus instability strip \citep{2003ApJ}. Our analysis revealed that J0543 and J0547 are both in the $\delta$ Scuti and the $\gamma$ Doradus instability strip.

To further investigate, we analyzed the pulsation behavior of the sample. The pulsation signals are relatively weak to detect. We removed the eclipse features by subtracting the fitted light curve template and then examined the residual light curves for evidence of pulsations. As a result, we found that J0039, J0547, and J1905, display significant pulsation signals. Below, we provide a detailed description of each case.

Our analysis for J0039 used SPOC-pipeline TESS light curves \citep{2020RNAAS...4..201C} from Sector 57. For J0547, we utilized existing Sector 59 data products from the TESS Quick-Look Pipeline \citep[QLP;][]{2022RNAAS...6..236K}. There is a nearby star within a distance of less than one pixel from J0547. We queried the {\it Gaia} DR3 for the $G$-band magnitudes of both stars: J0547 has a magnitude of 12.37, while the nearby star has a magnitude of 16.12. This indicates that the influence of the nearby star on our light curve can be considered negligible. Compared with J0039, J0547 exhibits a longer pulsation period of 0.38 days, which is atypical for normal $\delta$~Sct-type pulsations. We used the light curve from Kepler to analyze the properties of J1905. As illustrated in Figure~\ref{fig:pulsation}, the bottom panel highlights these signals. Consequently, the pulsation periods of these three objects range from 18 minutes to 8 hours. 

We find that J1905 is located outside the edges of instability strip in the HRD, as shown in Figure~\ref{fig:HRD}. This phenomenon is not unique, a significant fraction of objects with pulsation frequencies resembling those of $\delta$ Scuti-type variables have also been observed beyond the blue edge of the instability strip \citep{2024PASA...41...82K}. \citet{2017Liakos} analyzed a sample of binaries hosting $\delta$ Scuti stars and reported that approximately 15\% of these stars reside beyond the blue edge of the instability strip (see background stars in Figure~\ref{fig:HRD}). This suggests that pulsating stars may not be strictly confined to the instability strip. Although the pulsating component in the binary exhibits characteristics similar to those of normal pulsational stars, its pulsations might be influenced by the mass transfer and the tidal distortions \citep{2003ASPC..292..113M, 2006MNRAS.366.1289S}. To investigate potential correlations between orbital periods and pulsation properties in accretors, we overlaid a color bar representing orbital periods in Figure~\ref{fig:HRD}. However, no statistically significant trends were identified in either our sample or the background stars from \citet{2017Liakos}.

To determine the true nature of the pulsations in these three objects or the remaining sources, further analysis is required. Here, we only provide a preliminary indication of the pulsations without conducting more extensive research.

\section{Discussion} \label{sec:dis}

\begin{figure}
\includegraphics[width=0.5\textwidth]{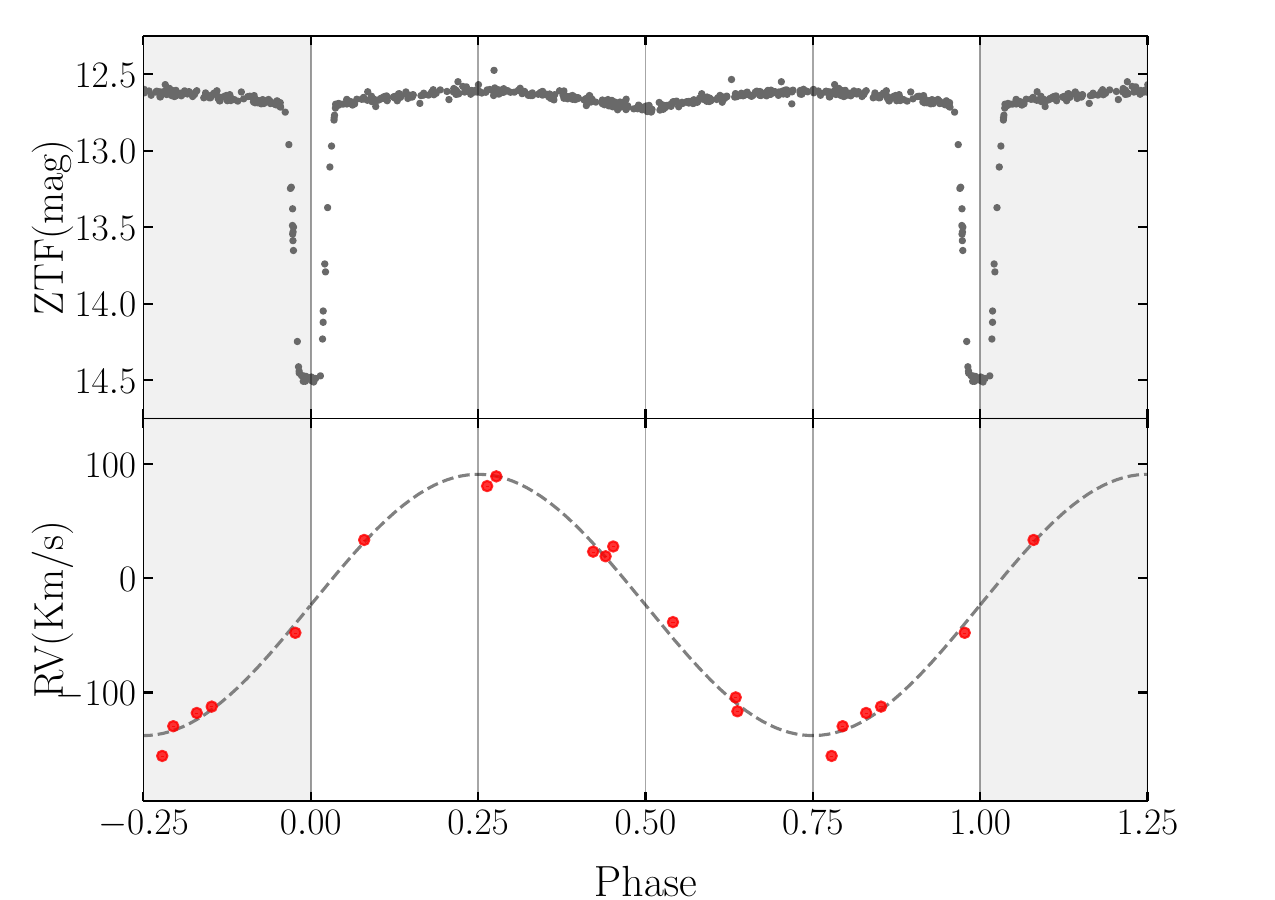}
\caption{The light curve and RV curve of J1905. The upper panel is the phase-folded light curve in the ZTF g band. The bottom panel is the phase-folded RV curve of the donor measured from the APOGEE spectra.
\label{fig:RV}}
\end{figure}

\subsection{Notes on individual objects} \label{subsec:individual-obj}
\subsubsection{J1905}
J1905 was observed with the APOGEE spectrograph, which has a spectral resolution of \(R \sim 22500\), over 14 epochs spanning from May 15, 2019, to November 3, 2020.

The cross-correlation function (CCF) profile exhibits a distinct single peak, indicating a single dominant component in the spectra. The donor's contribution to the flux in the APOGEE spectra is significantly greater than that of the accretor, leading to this result. Therefore, we utilized the APOGEE heliocentric radial velocities to represent the donor. We fit the folded radial velocity curve using a sinusoidal function form of a Keplerian orbit:\[ V_{\mathrm{r}}(t) = V_{0} + K_2 \sin{\left(\frac{2 \pi t}{P_{\mathrm{orb}}} + \phi_0\right)} \]where \(K_2\) is the semi-amplitude of the radial velocity curve, \(V_0\) is the system radial velocity of the binary, and \(\phi_0\) is the initial phase of the folding starting point. The fitted radial velocity curves are shown in Figure~\ref{fig:RV}, and the radial velocity semi-amplitude is \( K_2 = 113.8\pm 3.8\, \text{km s}^{-1} \). We calculated the mass function \(f(M_1)\) of this object using the equation below: \[ f(M_1) = \frac{(M_1 \sin i)^3}{(M_1 + M_2)^2} = \frac{P_{\rm orb} K_2^3}{2 \pi G} \]where \(i\) is the orbital inclination and \(G\) is the gravitational constant, resulting in \(f(M_1) = 1.93 \pm 0.18 \,M_\odot\). \citet{zhang2024photometric} used spectra obtained with the P200 telescope and LAMOST to measure the radial velocity curve for the donor of J1905. They determined a value of \( K_2 = 135.5 \pm 6.0 \, \text{km s}^{-1} \), which is similar to our result. 

Based on  the inclination, $M_2$ derived from the joint fitting, and the mass function, we calculated the accretor mass in J1905 to be $M_1 = 2.40 \pm 0.19 M_\odot$. This value shows discrepancy compared to the mass obtained from the joint fitting.
To investigate this inconsistency, we examined the light curve fitting results for this object and found that the shallower eclipse feature exhibits significant discrepancies between the data and the model (see Figure \ref{fig:lc}). These discrepancies may lead to deviations in the fitted inclination angle and orbital semi-major axis from their true values, ultimately contributing to the difference in the mass estimates.
\subsubsection{J0611 and J2107}
\label{subsubsec:discuss_J0611}

We used a 5\arcsec\ search radius in \Gaia DR3 to identify all sources in our sample. For J0611 and J2107, we discovered two distinct \Gaia\ Source IDs associated with each binary. Optical images from Pan-STARRS \citep{chambers2016pan} reveal that J2107 has a nearby star located just 2.66\arcsec\ away, while J0611 has a companion star at a separation of 1.94\arcsec.
Due to the substantial uncertainties in the parallax, proper motion from \Gaia\ DR3, we cannot conclusively determine whether these nearby stars are isolated field stars or components of the binary systems, potentially forming wide hierarchical triple systems.
However, the spectra of J0611 show the presence of the Na I D doublet, which may indicate the existence of a tertiary companion \citep{2024A&A...691A.260L}.

\subsection{Period-Mass relation of donors} \label{subsec:discuss_mass}

\begin{figure}
    \centering
    \includegraphics[width=\linewidth]{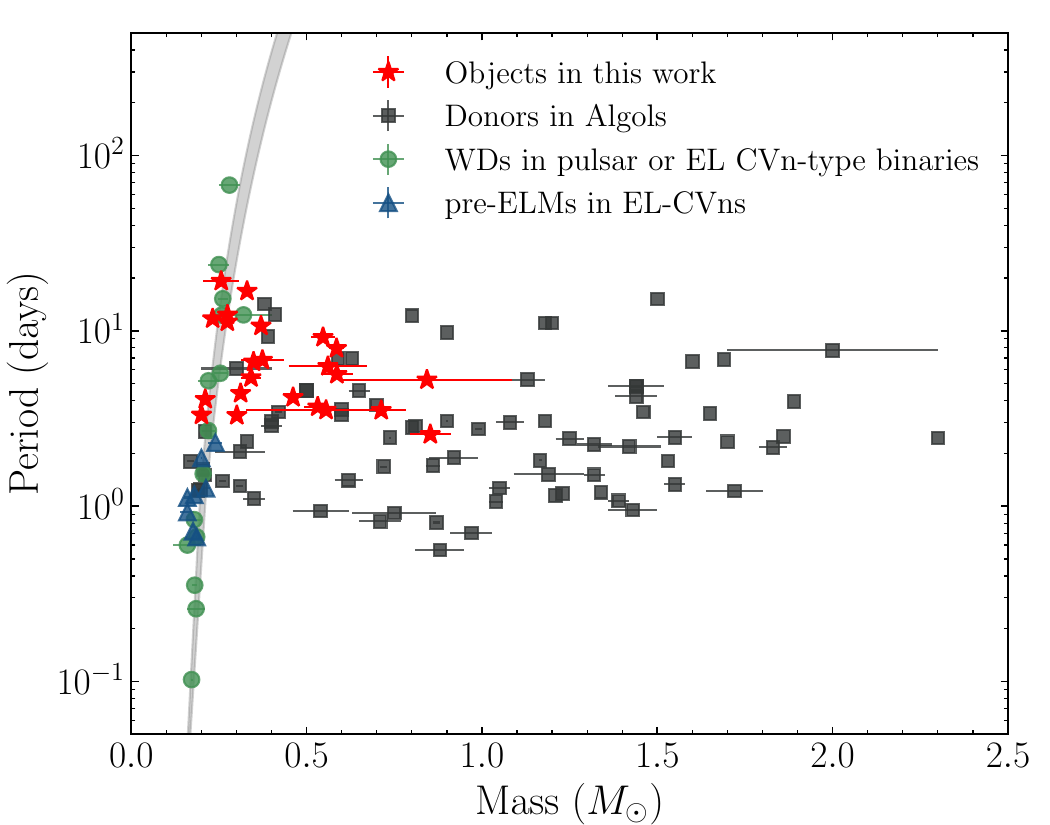}
    \caption{$M_\mathrm{WD}-P_\mathrm{orb}$ relation. The red star is the objects studied in this paper. The blue triangles are pre-ELMs in EL CVns. The green circles are helium WDs in binaries with companions of pulsars or A-type MS stars. The data of blue triangles and green circles are from the collection of \citet{zhang2022}. The black squares are donors in Algols, and the data are obtained from \citet{2010MNRAS.406.1071D,2011NewA...16..530L,2017AJ....154..216M,van2016accretion, 2022Ap&SS.367...22U,2024A&A...691A.260L}. The grey band is the $M_\mathrm{WD}-P_\mathrm{orb}$ relation provided by \citet{Tauris1999}. The upper and lower limits of the grey band correspond to metallicities of $Z = 0.001$ and $Z = 0.02$.}
    \label{fig:mwd-porb-relation}
\end{figure}

Statistical studies of binary systems composed of WDs and pulsars \citep[e.g.,][]{Tauris1999,van2005}, as well as EL-CVn type binaries \citep{Maxted2014}, reveal a significant correlation between the orbital period and the WD mass within these systems (see Figure \ref{fig:mwd-porb-relation}). This correlation is commonly referred to as the $M_\mathrm{WD}-P_\mathrm{orb}$ relation.

A prevalent explanation for this relation is that, during substantial expansion of a star, its core has already developed a degenerate helium (He) core. There is a notable correlation between the mass of the degenerate He core and the radius of the star's hydrogen envelope. When the donor's hydrogen envelope is nearly depleted, the correlation between the mass of He core and hydrogen envelope radius will lead to 
a strong correlation between the orbital period and the donor's mass, i.e., the $M_\mathrm{WD}-P_\mathrm{orb}$ relation \citep{Rappaport1995}.

In the binary systems studied in this work, the donors have undergone significant expansion and transferred most of their mass to their companions. According to the theoretical evolution of binaries, these donors will form helium cores. Therefore, we used the aforementioned relation to verify their evolutionary stages.
We plotted the orbital periods and donor masses of these binaries on the $M_\mathrm{WD}-P_\mathrm{orb}$ relation diagram (Figure \ref{fig:mwd-porb-relation}). The plot shows that approximately six donors follow the expected correlation, while the masses of other donors are notably higher than those predicted by the $M_\mathrm{WD}-P_\mathrm{orb}$ relation.

For donors with masses significantly exceeding the predictions of the $M_\mathrm{WD}-P_\mathrm{orb}$ relation, a plausible explanation is that these systems are still in the relatively early mass-transfer stages. As the hydrogen envelope continues burning and mass transfer proceeds, the donors will keep losing mass, leading to a further widening orbital separation. Eventually, as the mass-transfer process nears its completion, the orbital period and donor mass of these binary systems are expected to align with the predictions of the $M_\mathrm{WD}-P_\mathrm{orb}$ relation.

We also included a comparison with Algols studied by other researchers in the black squares of the $M_\mathrm{WD}-P_\mathrm{orb}$ relation in Figure \ref{fig:mwd-porb-relation}. Our systems are positioned closer to this relation than most of other Algols in the background. 
This may be a consequence of our selection criteria, which favor systems with large temperature ratios. As we discuss in Section \ref{subsec:why}, this criterion indirectly selects for donors with lower temperatures.
The ongoing mass transfer in these binaries leads to continuous expansion and cooling of the donor star \citep{zhou2018occultations, 2011A&A...528A..16V}.
Compared to the background sample, our donor stars exhibit lower temperatures and larger radii, 
suggesting that they are in a later stage of mass transfer, resulting in lower mass.

\subsection{Why our selection due to a sample of Algol-Type binaries with disks?\label{subsec:why}}

Our sample selection criteria focused on photometric and spectral characteristics without imposing constraints on system composition. Despite this lack of restriction, the parameters of all the objects demonstrate remarkable consistency.
Each system in our sample is an Algol-type binary consisting of an A-type star and a stripped bloated low-temperature star. We try to understand why our selection of objects yields such consistent results.
This uniformity may be attributed to the following three key observational features constrained.
 
The first key observational criterion is the substantial difference between the primary and secondary eclipse depths (eclipse depth ratio $>$ 20 in ZTF $g$ band). This extreme eclipse depth ratio indicates a significant temperature disparity between the two stars in the binary system, as exemplified by systems in our sample, such as J0641 with a temperature ratio of 2.24, and even the minimum temperature ratio is 1.55.
 
Given these ratios, the hotter star cannot plausibly correspond to F--type or cooler stars without requiring implausibly low temperatures for the cooler companion.
 
Regarding the observational selection of A-type primaries over hotter O- or B-type stars, we suspect that A-type stars are more common than brighter stars. Of course, we do not rule out there may be a more fundamental evolutionary explanation.

Additionally, the second observational criterion involving constraints from H$\alpha$ emission lines in the spectra effectively selects binary systems undergoing mass transfer \citep{carpenter1930u,atwood2012modeling,2020A&A...644A.121L}. Most of the systems we selected exhibit ellipsoidal characteristics in their light curves (see Figure~\ref{fig:lc}), further validating our hypotheses regarding stellar expansion and mass transfer.

The third observational criterion is the significant variability in the light curves ($\Delta \mathrm{mag} > 0.3$ in ZTF $g$ band). This limitation restrain the identification of systems where mass transfer occurs from the hotter primary stars to cooler secondary main-sequence stars.

In principle, other types of binary systems could also exhibit similar observational characteristics, such as binaries composed of a hot WD + M~dwarf. Theoretically, such systems could also show large eclipse difference and variability. However, among the 23 systems selected in our sample, we did not identify any binaries of this type. A possible explanation is that eclipsing WD + M~dwarf systems are intrinsically much rarer than Algol-type binaries. Given the limited sample size, their occurrence rate may be too low to appear in our selection. Additionally, the temporal sampling limitations and noise levels of ZTF may obscure detection of long-period eclipsing systems, making them easy to miss.

\section{Summary} \label{sec:summary}
We identified a sample of extreme eclipsing binary systems by selecting those characterized of large variability (\(\Delta \mathrm{mag} > 0.3\) in the \(g\) band) and significant differences between primary and secondary eclipses (eclipse depth ratio $>$ 20 in the \(g\) band). By cross-matching candidates selected from ZTF with the LAMOST spectroscopic survey, we obtained a sample of 23 Algol-type binaries.
Their orbital periods range from 2.57 to 19.21 days. We investigated the properties of these Algol-type binaries by jointly fitting their SEDs, spectra, and light curves, finding that the 23 sources are binaries consisting of highly stripped, low-luminosity subgiants and accreting A-type stars.
All of the samples have a bloated donor star with large radii from 2.5~\rsun\ to 8.9~\rsun\ and low temperatures (\(T_{\mathrm{eff}} \sim 4000 \, \mathrm{K}\)). The typical mass of the donors is \(M_{2} \sim 0.3 \,\msun\), showing that they have lost the majority of their masses. Both the large-amplitude ellipsoidal variability and the H$\alpha$ emission in the spectra indicate that the donors have filled their Roche lobes and the systems are in the stage of mass transfer.
The bright magnitudes in the optical band, low mass transfer rates, and stability make them excellent candidates for studying the accretion of low mass binaries without the disruptive effects of violent eruptions. We summarize the main characteristics of the samples below.
 
\begin{enumerate}

    \item The accretors appear to follow the main sequence on the color-magnitude diagram (CMD). 
    However, most of their luminosities are significantly lower than predicted by the mass-luminosity relation.
    This finding supports theoretical predictions that stars that have gained substantial mass from their companions undergo a phase of faint luminosity. Furthermore, it suggests that the accretors analyzed in this study have not yet returned to the main sequence. To better test the above statement, spectroscopic radial velocity observations are necessary.

    \item
    All of our binaries exhibit significant H\(\alpha\) emission, a crucial indicator of disk formation during mass transfer in low-mass accreting binaries. The H\(\alpha\) emission in our sample encompasses a wide variety of the four basic morphological types of Algol binaries. The relationship between the H\(\alpha\) emission line profiles and the orbital periods in our sample aligns with previous statistical results \citep{richards1999morphologies}, reinforcing the empirical understanding of this relationship. 

    \item 
    Nineteen of the accretors are located within the $\delta$~Sct instability strip, while two are situated within both the $\delta$~Sct and $\gamma$~Dor instability strip. Our analysis of residual light curves reveals that three of them exhibit distinct pulsation characteristics, with periods ranging from 18 minutes to 8 hours.
    Algol-type binaries with pulsating stars are ideal for studying stellar structure and evolution through asteroseismology and binary properties. The nature of the pulsations in these objects is worthy of further analysis in the subsequent work.
    
    \item 
    Six of the donors basically follow the $M_\mathrm{WD}-P_\mathrm{orb}$ relation, indicating that they should have entered the later stage of mass transfer. The masses of another sixteen donors are significantly higher than the expectation of $M_\mathrm{WD}-P_\mathrm{orb}$ relation, suggesting that they are in earlier stages of mass transfer. With the burning of the hydrogen shell and the process of mass transfer, the donors will lose more material and the binaries evolve towards wider orbits.

\end{enumerate}

\section{acknowledgments}
We thank Yang Huang and Mouyuan Sun for helpful discussion, and thank the anonymous referee for constructive suggestions that improved the paper. This work was supported by the National Key R\&D Program of China under grants 2023YFA1607901 and 2021YFA1600401, the National Natural Science Foundation of China under grants 12433007, 11925301, 12033006, 12221003, and 12263003. We acknowledge the science research grants from the China Manned Space Project.

We acknowledge the data provided by LAMOST. Guoshoujing Telescope (the Large Sky Area Multi-Object Fiber Spectroscopic Telescope LAMOST) is a National Major Scientific Project built by the Chinese Academy of Sciences. Funding for the project has been provided by the National Development and Reform Commission. LAMOST is operated and managed by the National Astronomical Observatories, Chinese Academy of Sciences.

We acknowledge the data provided by SDSS. 
SDSS-IV is managed by the 
Astrophysical Research Consortium 
for the Participating Institutions 
of the SDSS Collaboration including 
the Brazilian Participation Group, 
the Carnegie Institution for Science, 
Carnegie Mellon University, Center for 
Astrophysics | Harvard \& 
Smithsonian, the Chilean Participation 
Group, the French Participation Group, 
Instituto de Astrof\'isica de 
Canarias, The Johns Hopkins 
University, Kavli Institute for the 
Physics and Mathematics of the 
Universe (IPMU) / University of 
Tokyo, the Korean Participation Group, 
Lawrence Berkeley National Laboratory, 
Leibniz Institut f\"ur Astrophysik 
Potsdam (AIP),  Max-Planck-Institut 
f\"ur Astronomie (MPIA Heidelberg), 
Max-Planck-Institut f\"ur 
Astrophysik (MPA Garching), 
Max-Planck-Institut f\"ur 
Extraterrestrische Physik (MPE), 
National Astronomical Observatories of 
China, New Mexico State University, 
New York University, University of 
Notre Dame, Observat\'ario 
Nacional / MCTI, The Ohio State 
University, Pennsylvania State 
University, Shanghai 
Astronomical Observatory, United 
Kingdom Participation Group, 
Universidad Nacional Aut\'onoma 
de M\'exico, University of Arizona, 
University of Colorado Boulder, 
University of Oxford, University of 
Portsmouth, University of Utah, 
University of Virginia, University 
of Washington, University of 
Wisconsin, Vanderbilt University, 
and Yale University.

This paper also contains the data collected by ZTF, and TESS. ZTF is Supported by the National Science Foundation under Grants No. AST-1440341 and AST-2034437 and a collaboration including current partners Caltech, IPAC, the Oskar Klein Center at Stockholm University, the University of Maryland, University of California, Berkeley, the University of Wisconsin at Milwaukee, University of Warwick, Ruhr University, Cornell University, Northwestern University and Drexel University. Operations are conducted by COO, IPAC, and UW. 

This publication also uses datas from WISE, which is a joint project of the University of California, Los Angeles, and the Jet Propulsion Laboratory/California Institute of Technology, funded by the National Aeronautics and Space Administration. This publication also makes use of data products from NEOWISE, which is a project of the Jet Propulsion Laboratory/California Institute of Technology, funded by the Planetary Science Division of the National Aeronautics and Space Administration.

\software{Astropy \citep{2013astropy, 2018astropy}, lightkurve \citep{2018lightkurve}, emcee \citep{foreman2013emcee}, ELLC \citep{Maxted+16}, pyphot \citep{2025pyphot}, PyAstronomy \citep{2019PyAstronomy}, spectool \citep{spectool2025}, stellarSpecModel \citep{stellarSpecModel2025}, selenium \citep{selinium2022}}

\bibliography{ref}{}
\bibliographystyle{aasjournal}

\end{document}